\documentclass[aps,prb,twocolumn,superscriptaddress,nofootinbib,longbibliography]{revtex4-2}

\usepackage{amsmath}
\usepackage{amssymb}
\usepackage{graphicx}
\usepackage{bm}
\usepackage{hyperref}
\usepackage{xcolor}

\begin{document}

\title{Transient selection of competing order in frustrated spin Peierls systems \\ after a quench}

\author{Jing Zhou}
\affiliation{Institute of Physics, Chinese Academy of Sciences, Beijing 100190,China}
\affiliation{University of the Chinese Academy of Sciences, Beijing 100049, China}

\author{Yuan Wan}
\email{yuan.wan@iphy.ac.cn}
\affiliation{Institute of Physics, Chinese Academy of Sciences, Beijing 100190,China}
\affiliation{University of the Chinese Academy of Sciences, Beijing 100049, China}
\affiliation{Songshan Lake Materials Laboratory, Dongguan, Guangdong 523808, China}

\date{\today}

\begin{abstract}
We study theoretically the dynamics of frustrated spin Peierls systems after a quench from the paramagnetic state to the magnetically ordered state. By constructing and numerical simulating a minimal model, we show that it can exhibit a transient non-collinear magnetic order before settling in the equilibrium collinear magnetic order. The transient magnetic order is selected by the phonon fluctuations originated from the initial state. Our results reveal a mechanism for controlling the phases of matter by exploiting incoherent, nonequilibrium fluctuations.
\end{abstract}

\maketitle

The rise of ultrafast optics and spectroscopies opens a non-thermal pathway for controlling the phases of matter~\cite{Giannetti2016,Basov2017,Torre2021}. For instance, pumping the system with a short laser pulse can quench one long range order in favor of the other in a system with multiple competing orders, which makes it possible to guide the system to a desired yet thermodynamically metastable or even unstable state.

Frustrated magnets represent a prominent class of systems where competing orders appear naturally. A frustrated magnet hosts distinct magnetic orders that are accidentally degenerate in energy~\cite{Shender1996,Chalker2011}.  Quantum/thermal fluctuations, or chemical/structural disorders may lift this accidental degeneracy through the order by disorder (ObD) mechanism~\cite{Villain1980,Shender1982,Henley1989}, thereby stabilizing one magnetic order out of the many.

Among systems that host competing orders, frustrated magnets are unique in that their free energy landscape is shaped by fluctuations. This trait opens new possibilities for controlling their phases out of the equilibrium. While the much explored Floquet engineering employs periodic motion of normal modes~\cite{Mankowsky2016,Oka2019,Wan2017,Wan2018,Sun2024}, less attention is paid to \emph{incoherent} fluctuations. In this work, we reveal a mechanism by which transient magnetic orders emerge from such fluctuations in frustrated magnets coupled to soft phonons. After a quench from the paramagnetic state to magnetically ordered state, the system can develop a thermodynamically unstable, competing magnetic order owing to the nonequilibrium phonon fluctuations, before it settles in the equilibrium one.

In equilibrium, a frustrated magnet exhibits the Peierls instability in the presence of spin-phonon coupling ~\cite{Becca2002,Tchernyshyov2002a,Tchernyshyov2002b,Weber2005}. The instability arises because the system can release the frustration by spontaneously distorting the lattice. When the spin-orbit coupling is weak, the spin-Peierls instability selects a \emph{collinear} magnetic order among the degenerate ones.

We examine the evolution of the frustrated spin-Peierls systems after a quench from paramagnetic state to the magnetically ordered state. In typical pump-probe experiments on magnets with femtosecond laser pulses, the magnetic order melts rapidly, and then revives slowly as the system equilibrates~\cite{Wang2006,Kirikyuk2010,Koopmans2010,Torre2021}. We assume the optical pumping creates a paramagnetic state with an effective bath temperature $T_i$. When the pumping is off, the bath cools to a lower temperature $T_f$ over a relatively short time scale. We study the revival of the magnetic order as it adapts to the new bath temperature within the framework of time-dependent Ginzburg-Landau (TDGL) theory~\cite{Yusupov2010,Kung2013,Tagaras2019,Sun2020,Dolgirev2020a,Dolgirev2020b}, which exposes the essential ingredients for the physics discussed here. The microscopic mechanisms for the melting and revival of magnetic order are complex. It can nonetheless be phenomenologically modeled by the TDGL formalism as a first approximation. Our protocol is also equivalent to a temperature quench~\cite{Castelnovo2010,Hart2019}.

\begin{figure}
\includegraphics[width = 0.9\columnwidth]{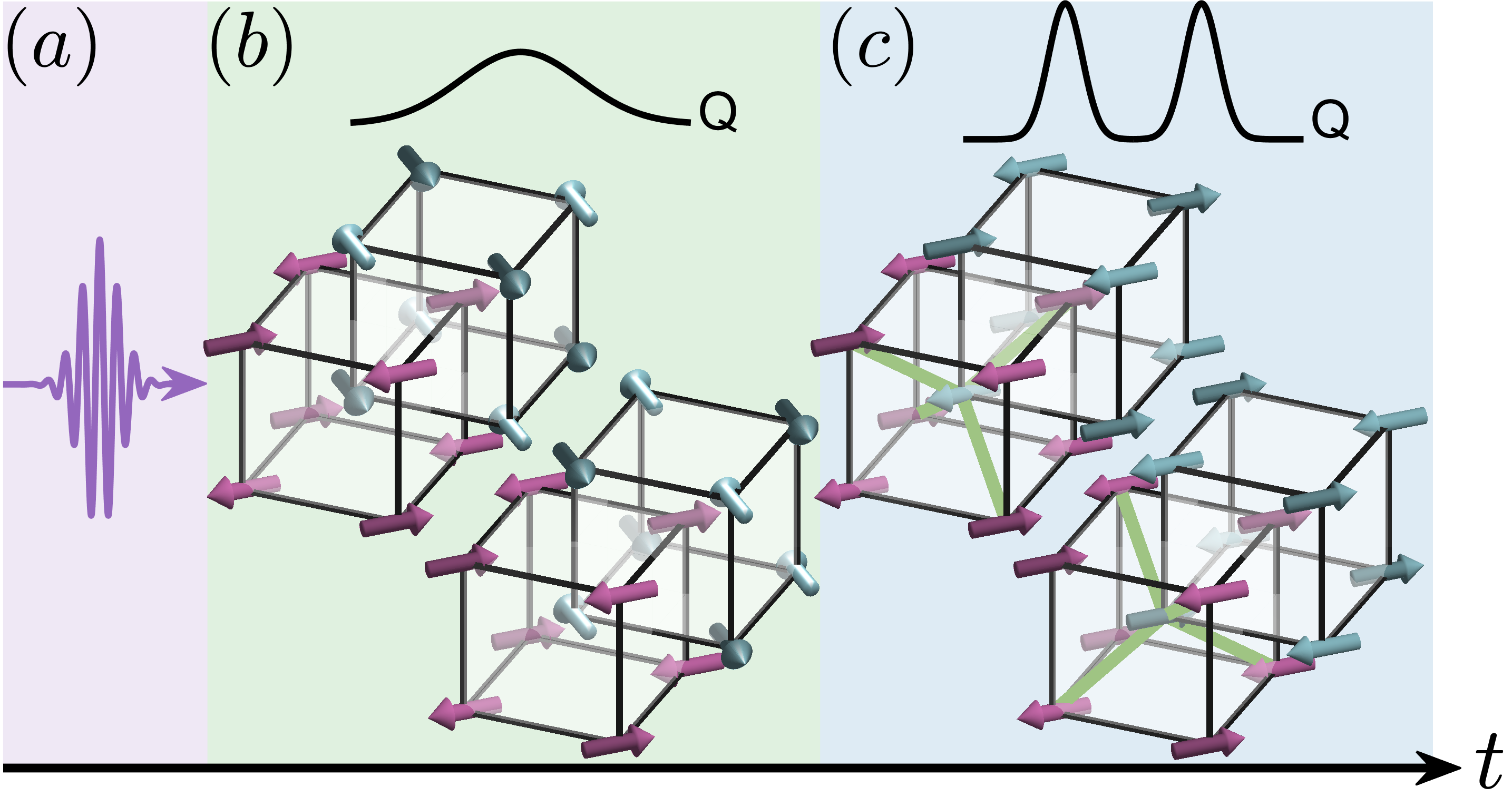}
\caption{Transient selection of competing magnetic order in a frustrated spin-Peierls system. (a) Paramagnetic state induced by optical pump. (b) Symmetry equivalent states of the transient non-collinear magnetic order, where the N\'{e}el orders in the two sublattices (magenta and cyan arrows) are orthogonal. Black curve sketches the distribution of phonon canonical coordinate $Q$. (c) Thermally stable collinear magnetic order. Exchange interaction are enhanced on green bonds due to spontaneous lattice distortion. The double peaked distribution of $Q$ reflects the two equilibrium positions.}
\label{fig:intro_v3}
\end{figure}

The phonon modes, described by their normal coordinates $Q$, are randomly distributed about $Q=0$ in the initial paramagnetic state. Meanwhile, when the phonons are soft, their dynamics can be much slower than that of the spins. The phonons generate an effective, quenched disorder for the frustrated spins through the spin-phonon coupling at the initial stage of the time evolution. The quenched disorder, through the ObD mechanism, selects the \emph{non-collinear} magnetic order~\cite{Henley1989} as opposed to the collinear magnetic order favored by the Peierls instability. As a result, the system must exhibit a tendency toward the thermally unstable non-collinear magnetic order before reverting to the collinear one (Fig.~\ref{fig:intro_v3}).

The dynamics of the system falls into the category of the phase ordering process~\cite{Bray1994,Puri2009,Cugliandolo2015}. As the initial state is symmetric, the system develop multiple domains in the thermodynamically limit. The transient competing order manifests itself as the dominant correlations, rather than true long range order, in the early time, and survive as a  subdominant short range order in the late time.

We expect that the above scenario is generic in that our arguments rely on the separation of time scales and the common features of the ObD physics. It can be viewed as a natural interpolation between two opposing order selection mechanisms in frustrated magnetism, namely the quenched disorder and the lattice distortion. In what follows, we construct a minimal model for frustrated spin-Peierls system, and demonstrate the transient magnetic order therein by a numerical simulation. 

We consider an easy-plane, body-centered cubic (BCC) lattice antiferromagnet  with both first neighbor ($J_1$)  and second neighbor ($J_2$) exchange interactions,whose ObD physics is among the simplest~\cite{Shender1982,Yildirim1996,Yildirim1999,Schmidt2002}. The BCC lattice comprises of two interpenetrating simple cubic sublattices, dubbed A and B, respectively. When $J_2>2J_1/3$, each sublattice hosts a N\'{e}el order, which are decoupled in the mean field limit (Fig.~\ref{fig:intro_v3}b\&c). 

We write down the symmetry constrained Landau free energy, $F = F_\mathrm{ex} +  F_\mathrm{s-ph}$~\cite{supp_mat}. $F_\mathrm{ex}$ describes the exchange interaction:
\begin{subequations}
\begin{align}
F_\mathrm{ex} = \frac{K}{2}\int [(\nabla \theta_A)^2 + (\nabla\theta_B)^2 + 2\kappa\nabla\theta_A\nabla\theta_B] d^3x.
\end{align}
$\theta_A$ and $\theta_B$ describes the direction of N\'{e}el order in sublattice A and B, respectively. $K,\kappa$ are spin stiffness. It possesses two $U(1)$ symmetries: The common rotation of $\theta_A$ and $\theta_B$ corresponds to the spin rotational symmetry. By contrast, the relative rotation reflects the accidental degeneracy due to the frustration. We have used the simplifying approximation that the magnitude of the N\'{e}el order parameters are fixed, which means the short-range order is not affected by the quench. This approximation can be removed by using the usual TDGL model, which yields qualitatively similar results~\cite{supp_mat}.

The spin-phonon part reads:
\begin{align}
F_\mathrm{s-ph} = \int [ -\lambda Q \cos(\theta_A - \theta_B) + \frac{\epsilon}{2} Q^2 ] d^3x.
\end{align}
\end{subequations}
$Q$ is the normal coordinate of an Einstein phonon. The spin-phonon coupling constant $\lambda>0$. Microscopically, it describes the modulation of the $J_1$ interactions by the displacement in the non-magnetic ions (not in BCC lattice) that mediate the superexchange (Fig.~\ref{fig:intro_v3}c). $\epsilon>0$ is the elastic constant. 

At equilibrium, minimizing $F$ yields the ground states, $\theta_A = \theta_B$, $Q = \lambda/\epsilon$, or $\theta_A = \theta_B+\pi$, $Q = -\lambda/\epsilon$. These solutions describe the collinear magnetic orders, accompanied by a spontaneous distortion of the lattice  (Fig.~\ref{fig:intro_v3}c).

We endow the system with model A dynamics~\cite{Hohenberg1977}: 
\begin{align}
\dot{\theta}_{A,B} = -\Gamma_\mathrm{s}\frac{\delta F}{\delta\theta_{A,B}} + \xi_{A,B};
\quad
\dot{Q} = -\Gamma_\mathrm{ph}\frac{\delta F}{\delta Q} + \xi_{Q}.
\end{align} 
$\Gamma_\mathrm{s}$ and $\Gamma_\mathrm{ph}$ are respectively the kinetic coefficients of the spins and the phonons. $\xi_{A,B,Q}$ are independent Gaussian white noise fields, whose covariance are $\langle \xi_{\alpha}(x,t)\xi_{\alpha}(x',t')\rangle =  2\Gamma_\alpha k_BT\delta^{(3)}(x-y)\delta(t-t')$. $\Gamma_\alpha = \Gamma_\mathrm{s} (\Gamma_\mathrm{Q})$ when $\alpha=A,B(Q)$. $T$ is the bath temperature.

We make a couple of extra steps to simplify the above model further. We define a pair of fields, $\psi = (\theta_A+\theta_B)/2$, and $\theta = \theta_A-\theta_B$. The equation of motion for $\psi$ and $\theta$ are then decoupled. $\psi$ is the order parameter corresponding to the spontaneous breaking of the spin rotational symmetry. $\theta$ parametrizes the continuous family of degenerate magnetic orders; the two N\'{e}el vectors are collinear when $\theta = 0$ or $\pi$, whereas they are orthogonal when $\theta = \pm \pi/2$. Since the focus is the selection of magnetic orders, we drop $\psi$ for now. The resulted minimal model is silent on the magnitude of the N\'{e}el order. A direct analysis that treats $\psi$ and $\theta$ on equal footing yields similar behavior in terms of magnetic order selection~\cite{supp_mat}.

We discretize the model on a simple cubic lattice for numerical simulation. After rescaling~\cite{supp_mat}:
\begin{subequations} \label{eq:minimal_model}
\begin{align}
\dot{\theta}_i = -\frac{\partial \widetilde{F}}{\partial\theta_i} + \nu_i; \quad
\dot{\widetilde{Q}}_i = -\frac{1}{\tau}\frac{\partial \widetilde{F}}{\partial \widetilde{Q}_i}+\eta_i,
\end{align}
with the dimensionless free energy function:
\begin{align}
\widetilde{F} = -\sum_{\langle ij\rangle}\cos(\theta_i-\theta_j)  -g\sum_i \widetilde{Q}_i\cos\theta_i + \sum_i \frac{\widetilde{Q}^2_i}{2}.
\end{align}
\end{subequations}
We define dimensionless time variable $\widetilde{t} = t/\tau_\theta$, with $\tau_\theta = a^2/(2K_\theta \Gamma_s)$ being the relaxation time for lattice scale spin fluctuations. $a$ is lattice cutoff. $K_\theta = K(1-\kappa)/2$ is the stiffness for $\theta$. $\tau = 2\Gamma_s K_\theta/(a^2\Gamma_\mathrm{ph}\epsilon)$ is the relaxation time of the phonons measured in $\tau_\theta$. Crucially, $\tau\gg1$ when the phonons are soft ($\epsilon\to0$). $g=\sqrt{a^2/\epsilon K_\theta}\lambda$ and $\widetilde{Q} = \sqrt{\epsilon a^2/K_\theta} Q$ are respectively the dimensionless spin-phonon coupling and phonon coordinate. The Gaussian white noises $\nu_i$ and $\eta_i$ have the covariance $\langle \nu_i(\widetilde{t})\nu_j(\widetilde{t}')\rangle = 2\widetilde{T}\delta_{ij}\delta(\widetilde{t}-\widetilde{t}')$, and $\langle \eta_i(\widetilde{t}) \eta_j(\widetilde{t}'))\rangle = 2(\widetilde{T}/\tau) \delta_{ij}\delta(\widetilde{t}-\widetilde{t}')$. The dimensionless bath temperature $\widetilde{T} = k_BT/(K_\theta a)$.

We estimate $g \sim \sqrt{\delta J/J}$, where $J$ is the spin exchange constant and $\delta J$ its modulation in the spin-Peierls phase. Setting $\delta J/J\sim 10^{-2}$ results in $g\sim 0.1$~\cite{Jaubert2024}. $\tau_\theta\sim 1$  to $10$ps estimated from spin wave lifetime~\cite{Bayrakci2006,Gilmore2007}. Phonon relaxation time varies greatly ~\cite{Sushkov2005,AkuLeh2005}, yielding a ratio $\tau$ in the range of 1 to hundreds. The effective model parameters may depend on temperature, which can affect quantitative but not qualitative behavior of the model.

Eq.~\eqref{eq:minimal_model} is the starting point of the ensuing analysis. We set the initial temperature $\widetilde{T}_i = 3$. As $\widetilde{T}_i \gg $ the energy scale for spin exchange and spin-phonon coupling, we approximate the initial state by an uncorrelated state to reduce the computational cost, where $\theta_i$ are drawn independently from a uniform distribution over $[0,2\pi)$, and $\widetilde{Q}_i$ from a Gaussian distribution with $\langle \widetilde{Q}^2_i \rangle = \widetilde{T}_i$ as per the equipartition theorem. Simulations with correlated initial state yield quantitatively similar results~\cite{supp_mat}. Changing $\widetilde{T}_i$ doesn't alter the qualitative behavior provided it is far above the ordering temperature $\approx 2$.

We begin with a qualitative analysis of Eq.~\eqref{eq:minimal_model} in a couple of limits. In the limit of $\tau\to 0$, the phonons response immediately to the spins; therefore, we may set $\widetilde{Q}$ to its instantaneous equilibrium value, $\overline{Q}_i = g\cos\theta_i$. Substituting $\widetilde{Q}_i$ by $\overline{Q}_i$ in the equation of motion for $\theta$, we obtain an effective Landau energy for the spins, $\widetilde{F} = -\sum_{\langle ij\rangle}\cos(\theta_i-\theta_j) -g^2/2 \sum_i \cos^2\theta_i$. The effective anisotropy term generated by the phonons lifts the accidental $U(1)$ degeneracy, and favors the collinear magnetic orders, namely $\theta = 0$ or $\pi$. Therefore, the system exhibits the equilibrium magnetic order after the quench.

In the opposite limit of $\tau\to\infty$, $\widetilde{Q}$ are frozen to their initial states $Q^{(0)}$. The frozen phonons give rise to a random magnetic field in the $x$ direction, which favors the non-collinear states ($\theta = \pm\pi/2$)~\cite{Aharony1978,Minchau1985,Feldman1998}. To see this, we consider a single domain of size $\xi \gg1$. Within the domain, the Landau energy has relaxed to its minimum. We write $\theta_i = \theta+\delta\theta_i$, where $\theta$ is the average orientation of the domain, and $\delta\theta_i$ is the local variation due to random field. When $g\ll1$, expanding $\widetilde{F}$ to the second order in $\delta\theta_i$ yields $\widetilde{F} = 1/2\sum_{\langle ij\rangle} (\delta\theta_i-\delta\theta_j)^2 + g \sin\theta \sum_i  Q^{(0)}_i\delta\theta_i$. Minimizing $\widetilde{F}$ with respect to $\delta\theta_i$, and average over $Q^{(0)}$, we obtain $\widetilde{F} = -\alpha \widetilde{T}_i  \xi^3g^2\sin^2\theta/2$. Here, $\alpha \approx 0.25$ is a constant. Thus, the frozen phonons generate an effective anisotropy for $\theta$, which favors $\theta = \pm \pi/2$, namely the states where the two N\'{e}el orders are orthogonal.

Having understood both limits, we consider the case with finite $\tau$. We drop the noise for simplicity. Solving the equation of motion of $\widetilde{Q}$, and substitute the solution into that of $\theta$, we obtain $\dot\theta_i = -\sum_{j\in N_i} \sin(\theta_i-\theta_j) - g Q^{0}_i  e^{-\widetilde{t}/\tau} \sin\theta_i- g^2/\tau \int^{\widetilde{t}}_0 e^{-\frac{\widetilde{t}-s}{\tau}}\sin\theta_i(\widetilde{t}) \cos\theta_i(s) ds$. The first summation on the right hand side is over all nearest neighbors. The second term is the random field $\parallel x$ due to the initial fluctuations. According to the preceding discussion, it generates an anisotropy that favors the non-collinear magnetic order. This anisotropy disappears after $\widetilde{t}>\tau$. Meanwhile, the third term is a retarded self-interaction mediated by phonons. For the dynamics over time scales $\gg\tau$, the Markov approximation reduces it to $-g^2\sin\theta(t)\cos\theta(t)$. This term selects the collinear order similar to the $\tau\to0$ limit. We conclude that the non-collinear magnetic order develops over the time window $\widetilde{t}<\tau$, and gives way to the collinear one when $\widetilde{t}>\tau$.

\begin{figure}
\includegraphics[width=\columnwidth]{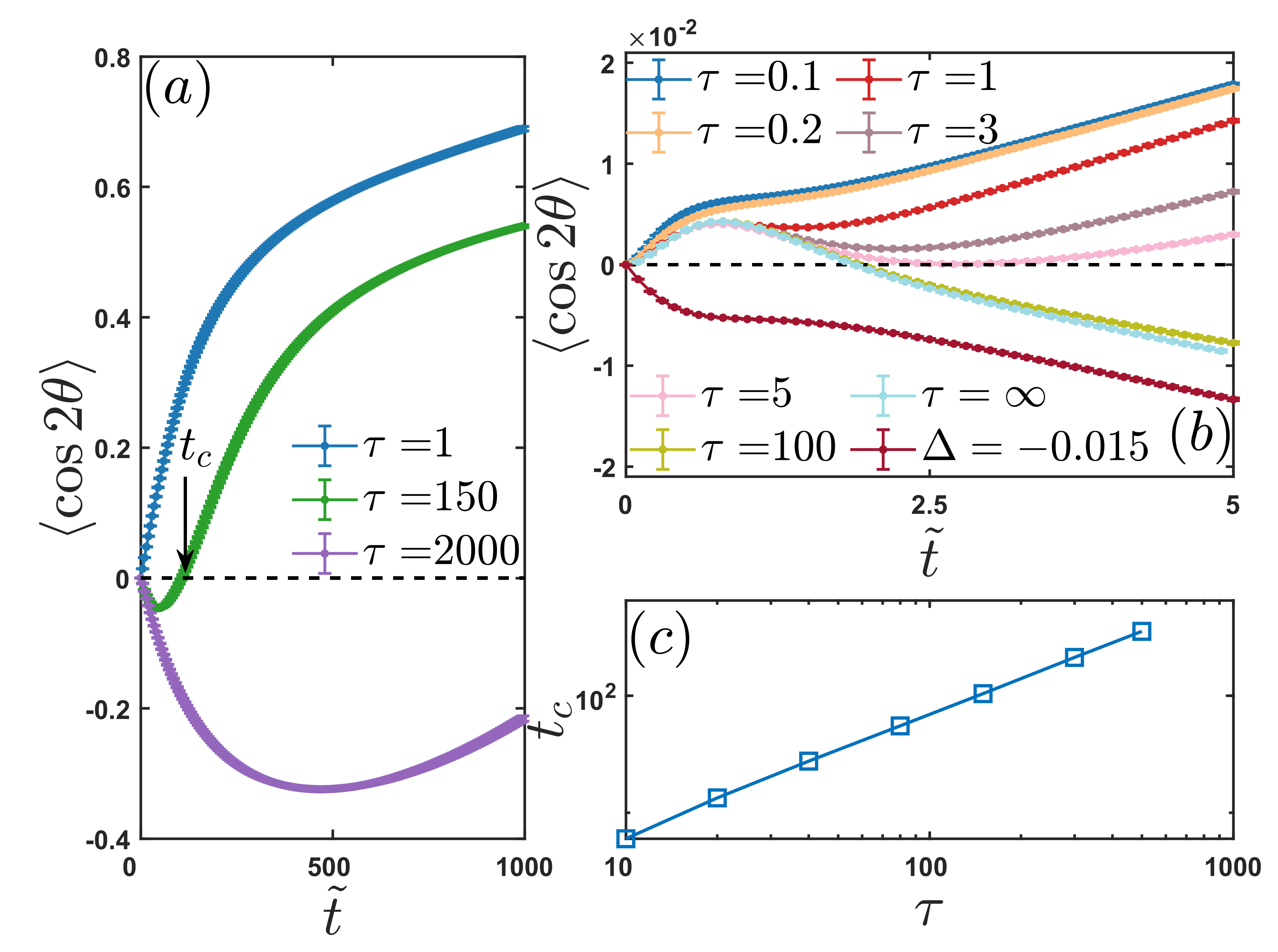}
\caption{(a) The local spin anisotropy, $\langle\cos(2\theta)\rangle$, as a function of time $t$ for various phonon relaxation times $\tau$. The system size $L=120$. Increasing $L$ doesn't yield discernible changes. (b) Early time behavior of the local spin anisotropy for various values of $\tau$ and a model with added anisotropy $-|\Delta|\sin^2\theta$. (c) Linear scaling between the spin reorientation time $t_c$ and the phonon relaxation time $\tau$.}
\label{fig:single_site} 
\end{figure}

We put this picture to test by numerically solving Eq.~\eqref{eq:minimal_model}. We use a $L\times L\times L$ lattice subject to periodic boundary conditions, with the maximal size $L=120$. We choose the representative model parameters $g = 0.2$ and $\widetilde{T}_f=0.1$. The results do not show qualitative change so long as $g \ll 1$ and $\widetilde{T}_f$ is far below the ordering temperature. We solve the Langevin equation using the Euler method with time step $dt = 10^{-2}$. All results are obtained by averaging over 840 runs.

We characterize the magnetic order selection by the observable, $\langle \cos(2\theta) \rangle =\sum_{i} \cos(2\theta_i)/L^3$. $\langle\cos(2\theta)\rangle = 1$ if $\theta_i = 0,\pi$ $\forall i$, and $-1$ if $\theta_i = \pm\pi/2$. Thus, this quantity measures the \emph{local} anisotropy. Fig.~\ref{fig:single_site}a shows the numerically calculated $\langle \cos(2\theta)\rangle$ as a function of time. For $\tau = 1$, it starts from 0 and increases in time, reflecting the building up of the \emph{local} collinear magnetic order. By contrast, for $\tau = 2000$, $\langle\cos(2\theta)\rangle$ is negative within the simulated time window, indicating the selection of the local non-collinear magnetic order. Its magnitude reaches the maximum at $\widetilde{t}\sim 300$. The non-monotonic behavior is due to the fact that the spin anisotropy generated by the phonon fluctuations is slowly decreasing in time. 

More interesting is the case with $\tau = 150$, where $\langle\cos(2\theta)\rangle$ changes sign at $t_c \approx 104$. This sign change is the manifestation of the change in the local magnetic order. In analogy with the spin reorientation transition in equilibrium~\cite{Horner1968}, we dub $t_c$ the spin reorientation time. 

It is natural to ask if observing the spin reorientation requires a minimum $\tau$. To this end, we examine the early time behavior of $\langle\cos2\theta\rangle$ (Fig.~\ref{fig:single_site}b). It starts out being positive at the initial stage for all values of $\tau$. For $\tau=1$, it remains positive throughout. On the other hand, for larger $\tau$, it crosses to the negative side and then changes its sign again at $t_c$. This sets a threshold at $\tau \approx 5$, below which the transient selection of the local non-collinear magnetic order does not occur.

The threshold stems from the fact that it takes a finite amount of time for the ObD phenomenon to establish. Even for $\tau\to\infty$, $\langle\cos2\theta\rangle$ starts out being positive, and then becomes negative at $\widetilde{t}\approx2$. We heuristically understand this behavior as follows. The initial dynamics are dominated by the random field $\parallel x$, which gives rise to a small but positive $\langle\cos2\theta\rangle$. The selection effect appears only after the exchange interaction generates spin correlations across a few lattice spacings. We stress that this behavior is a feature of the ObD physics; the sign change doesn't occur if the random field is replaced by an anisotropy term $-|\Delta|\sin^2\theta$ (Fig.~\ref{fig:single_site}b).

Finally, for $\tau$ above the threshold, we find a linear scaling between the spin reorientation time $t_c$ and the phonon relaxation time $\tau$ (Fig.~\ref{fig:single_site}c), in agreement with the picture from the qualitative analysis.

\begin{figure}
\includegraphics[width=\columnwidth]{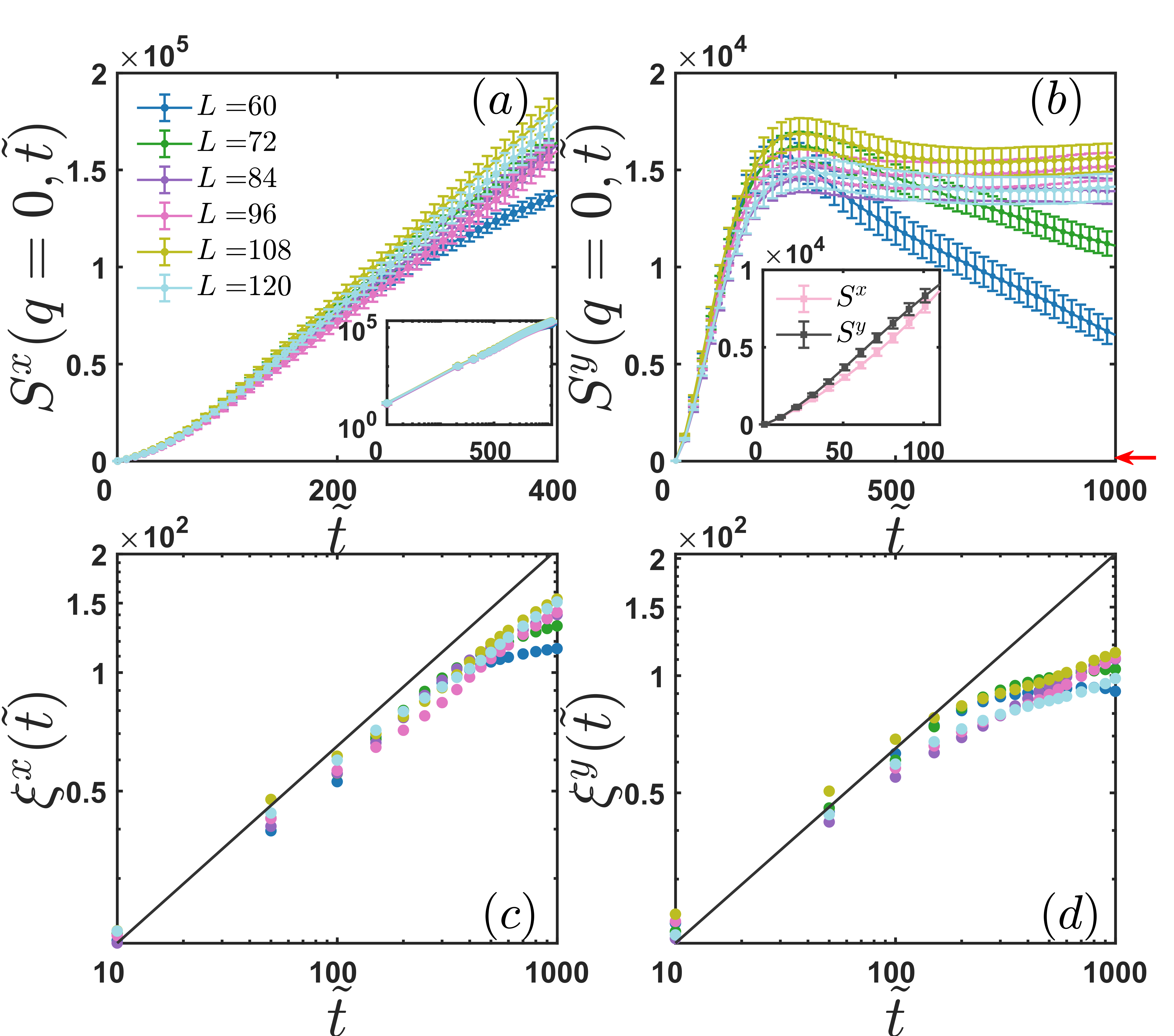}
\caption{(a) Structure factor $S^x(q=0)$ as a function of time for various system sizes. Phonon relaxation time $\tau = 150$. The inset shows the same data on logarithmic scale. The solid line denotes the $t^{3/2}$ scaling. (b) $S^y(q=0)$ as a function of time for various system sizes. Red arrow marks the equilibrium value. The inset shows a comparison between $S^x$ and $S^y$ for $L=120$ at early times. (c)(d) Correlation length $\xi^x$ and $\xi^y$ for the collinear and non-collinear magnetic order. The solid line denotes the $t^{1/2}$ scaling.}
\label{fig:tau_150} 
\end{figure}

The observable $\langle\cos2\theta\rangle$ is designed to probe  local magnetic order. We now investigate the non-local spin correlations by measuring the time-dependent structure factor, $S^{a}(\mathbf{q},\widetilde{t}) = \sum_{ij}\langle n^a_i(\widetilde{t}) n^a_j(\widetilde{t}) \rangle e^{-i\mathbf{q}\cdot (\mathbf{r}_{i} - \mathbf{r}_j)}/L^3$. $a = x,y$. $n^x_i = \cos\theta_i$, and $n^y_i = \sin\theta_i$. $\langle n^{x,y}_i\rangle = 0$ by symmetry. $S^x$ ($S^y$) probes the spatial correlation of the (non-)collinear magnetic order. 

Fig.~\ref{fig:tau_150}a shows $S^x(q=0)$, the height of the Bragg peak for the collinear magnetic order, as a function time. For prototypical phase ordering process such as that of the Ising model, the Bragg peak height exhibits a $\widetilde{t}^{3/2}$ law, reflecting the domain growth~\cite{Bray1994}. The domain size grows as $\xi \sim \widetilde{t}^{1/2}$. As the height is proportional to the domain volume, it grows as $\xi^3 \sim \widetilde{t}^{3/2}$. Here, $S^x(q=0)$ exhibits an approximate $\widetilde{t}^{3/2}$ law (Fig.~\ref{fig:tau_150}a, inset).

We extract a correlation length $\xi^x$ from $S^x(\mathbf{q})$ by measuring the half width at half maximum $q_0$ along the $(100)$ direction and using $\xi^x = 2\pi/q_0$~\cite{supp_mat}. We observe a deviation from the $\widetilde{t}^{1/2}$ growth law. At late time, the dynamics of the model is mapped to a three-dimensional XY model with easy axis anisotropy, which is in the same universality class as the Ising model. The deviation is likely due to uncertainties in measuring the correlation length at late time~\cite{supp_mat}.

We then turn to the competing non-collinear magnetic order. Fig.~\ref{fig:tau_150}b shows the evolution of $S^y(q=0)$. Crucially, $S^y(q=0)$ is larger than that of $S^x(q=0)$ at early time (Fig.~\ref{fig:tau_150}b, inset), demonstrating that the non-collinear magnetic order is the dominant correlation in the transient regime. The time when the two are equal is slightly shorter but comparable to that of $t_c$.

For small system size, the system is taken over by a single domain of collinear state with $\theta=0$ or $\pi$ at late time. We expect $S^y(q=0)$ will eventually decrease to its thermal equilibrium value (Fig.~\ref{fig:tau_150}b, red arrow). As $L$ increases, the slope of decrease is suppressed, suggesting it plateaus within numerically accessible time. The large non-collinear spin correlation compared to equilibrium is the relic of the transient selection of the competing order: The late time spin configurations comprise of many domains of collinear states. The non-collinear states survive near the domain walls~\cite{supp_mat}. It takes exceedingly long time for the non-collinear state to disappear due to the slow motion of the domain walls. 

The fact that the non-collinear magnetic order survives as a subdominant short-range order is mirrored in the correlation length (Fig.~\ref{fig:tau_150}d). After an initial approximate $\widetilde{t}^{1/2}$ growth, $\xi^y$ exhibits a kink near the spin reorientation time $t_c$ and continues to grow with much smaller slope.

\begin{figure}
\includegraphics[width=\columnwidth]{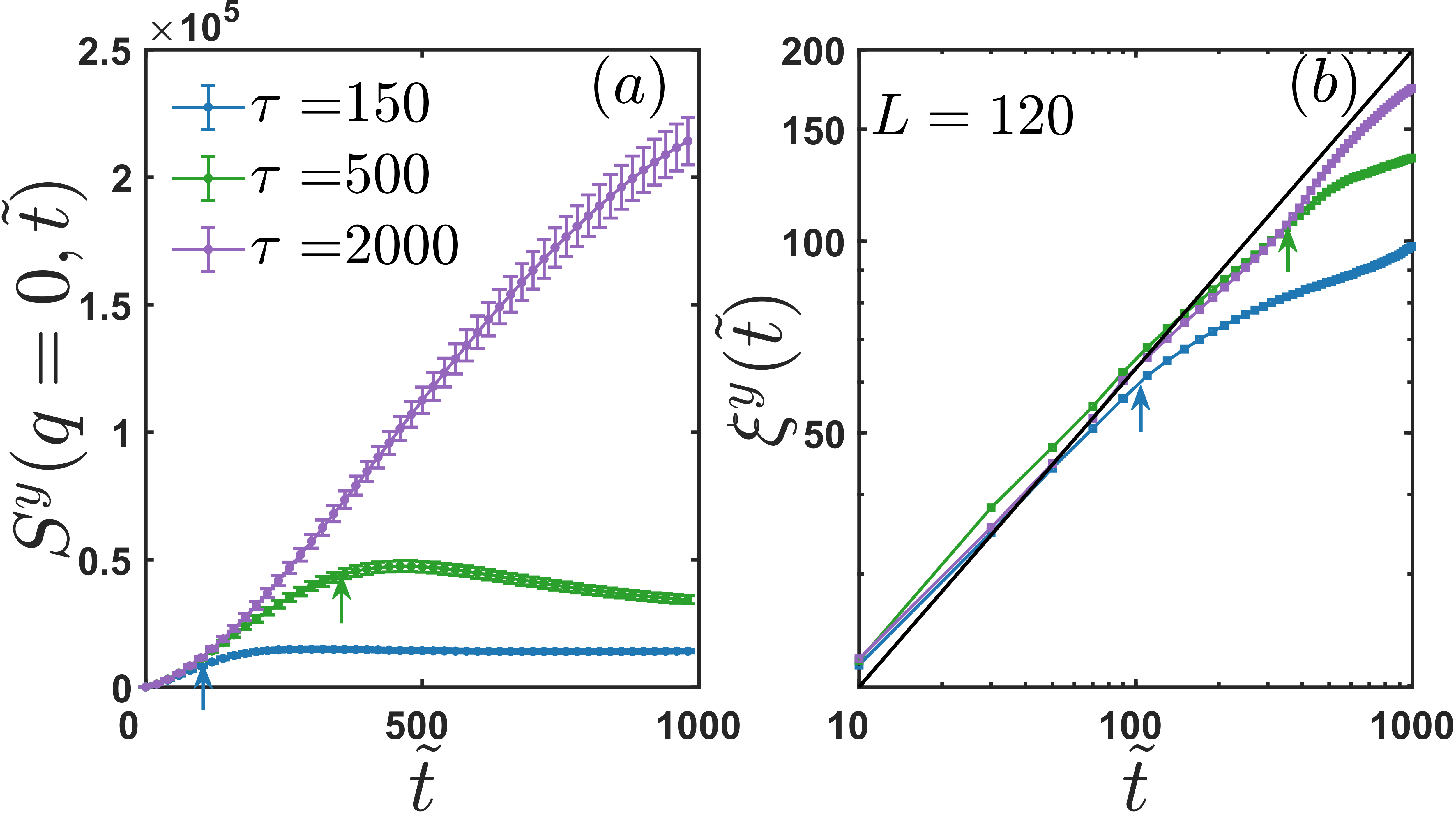}
\caption{ (a) Structure factor $S^y(q=0)$ as a function time for a few representative values of the phonon relaxation time $\tau$. $L=120$. (b) Correlation length $\xi^y$ as a function of time. Solid line marks the $t^{1/2}$ scaling. Arrows mark the spin reorientation time $t_c$.}
\label{fig:sy_scan_tau}
\end{figure}

We now compare the evolution of non-collinear magnetic order for different phonon relaxation time ratios $\tau$ (Fig.~\ref{fig:sy_scan_tau}). The overall magnitude of $S^y(q=0)$ and $\xi^y$ increases with $\tau$. The correlation length $\xi^y$ for $\tau=150$ and $\tau = 500$ both show an initial approximate $\widetilde{t}^{1/2}$ growth, followed by a slower growth after the spin reorientation time $t_c$. The data for $\tau = 2000$ only show the approximate $\widetilde{t}^{1/2}$ growth within the simulated time window.

To conclude, we have demonstrated that a frustrated spin-Peierls system can exhibit a transient, competing magnetic order owing to the nonequilibrium, incoherent phonon fluctuations. While the analysis is carried out for a BCC lattice model for simplicity, whose material incarnations are rare, we expect the mechanism applies to systems with other lattice geometries. Aside from its square lattice cousin~\cite{Melzi2001,Bombardi2004}, preliminary analysis of triangular and pyrochlore lattice models, motivated by  delafossites~\cite{Kimura2006,Ye2006,Ye2007,Plumer2007,Wang2008,Quirion2009,Damay2009,Vecchini2010,Giot2007,Zorko2008,Zorko2014} and chromate spinels~\cite{Lee2000,Sushkov2005,Bergman2006,Matsuda2007,Matsuda2010,Takagi2011,Tsurkan2021,Jaubert2024}, show that the non-equilibrium phonon fluctuations favor magnetic orders that are distinct from those by lattice distortion~\cite{supp_mat}. 

The order by quantum/thermal disorder can be incorporated into our model by adding a term $-\Delta \cos^2(\theta_A-\theta_B)$~\cite{supp_mat}, which suppresses the non-collinear state. In the quantum case, $\Delta \propto 1/S$, where $S$ is the quantum spin number. It is likely modest in delafossites and chromate spinels for their large $S$. In the thermal case, $\Delta \propto \widetilde{T}_f$, tunable with experimental conditions. The spatial correlation of phonon fluctuations, omitted in the present analysis, do not alter the transient selection of competing order provided that the correlation is short-ranged.

Our work points to a couple of directions for further research. The setup is mapped to a peculiar phase ordering process, namely an XY model whose easy-axis anisotropy changes sign in time. An analysis of this case will shed light on how the spatial correlation of the transient order evolves in time. Our model can also be extended to the case of underdamped phonons. Preliminary numerical simulation shows, after a quench, they produce qualitative similar behavior as underdamped phonons in the slow relaxation regime~\cite{supp_mat}. The impact of nonequilibrium fluctuations on competing phases unveiled in this work may be fruitfully explored in the context of intertwined and vestigial orders~\cite{Fradkin2015,Fernandes2019,Sun2020,Grandi2024}.

\begin{acknowledgments}
This work is supported by the National Key R\&D Program of China (Grant No.2022YFA1403800, 2024YFA1408700), National Natural Science Foundation of China (Grant No.12250008, 11974396, and 12188101), and by the CAS Project for Young Scientists in Basic Research (Grant No. YSBR-059). 
\end{acknowledgments}

The data that support the findings of this article are openly available~\cite{dataset}.

\bibliography{obd_quench}

\appendix
\onecolumngrid

\section{Construction of the Landau free energy}

We parametrize the N\'{e}el orders on the sublattice A and B by $\theta_A$ and $\theta_B$, respectively. We tabulate the transformation properties of these two variables under relevant symmetry operations of the system in Table~\ref{tab:symmetry}.

\begin{table}
\begin{tabular}{cc}
\hline\hline
Symmetry operations & Transformation \\
\hline
Spin rotation & $\theta_A \to \theta_A + \alpha;\, \theta_B \to \theta_B + \alpha$ \\
Time reversal & $\theta_A \to \theta_A+\pi;\, \theta_B \to \theta_B + \pi$ \\
Inversion w.r.t. an A site & $\theta_A \to \theta_A ;\, \theta_B \to \theta_B + \pi$ \\
Inversion w.r.t. an B site & $\theta_A \to \theta_A+\pi;\, \theta_B \to \theta_B$ \\
Half translation along $[111]$ & $\theta_B \to \theta_A$, $\theta_A\to \theta_B+\pi$ \\
\hline\hline
\end{tabular}
\caption{Symmetry transformations on $\theta_A$ and $\theta_B$.}
\label{tab:symmetry}
\end{table}

We are now ready to construct the symmetry constrained Landau energy. For the exchange interaction, we seek terms that are quadratic in the spatial gradient of $\theta_A,\theta_B$. It is easy to see that the symmetry allowed terms are $(\nabla\theta_A)^2+(\nabla\theta_B)^2$ and $\nabla\theta_A\nabla\theta_B$. We thus find:
\begin{align}
F_\mathrm{ex} = \frac{K}{2} \int [(\nabla\theta_A)^2+(\nabla\theta_B)^2+2\kappa \nabla\theta_A\nabla\theta_B]d^3x.
\end{align}
Imposing more symmetry requirements do not change the above form.

We recall that, microscopically, the spin-phonon coupling is of the form $Q\mathbf{S}_i\cdot\mathbf{S}_j$. In this work, we consider the so-called bond phonon model for the sake of simplicity: we assume that the displacement in the anions that mediate the super-exchange modulates the strength of exchange interaction, and that the modulation on these bonds are independent from each other. The modulation pattern shown in the main text is uniform within each simple cubic sub-lattice but odd under the half-translation along $[111]$. This pattern corresponds to the condensation of a phonon mode $Q$ that is at the high symmetry point $H = [100]$ of the first Brillouin zone. The symmetry of this phonon mode is classified by the little co-group $O_h$. It is easy to see that this mode carries irreducible representation $A_{2u}$.

After coarse-graining, the spin bilinear $\mathbf{S}_i\cdot\mathbf{S}_j$ would give rise to terms of the form $\cos(m\theta_A+n\theta_B)$ or $\sin(m\theta_A+n\theta_B)$. The spin rotational symmetry rules out terms with $m\neq -n$. We thus only need to consider $\cos(m\theta_A-m\theta_B)$ and $\sin(m\theta_A-m\theta_B)$. The symmetry requires that this term is odd under half-translation and carries the $A_{2u}$ representation. The half translation symmetry rules out the sine harmonics. As for the cosine harmonics, the lowest order harmonics is $\cos(\theta_A-\theta_B)$, and shows the same transformation properties as $Q$. We thus obtain the spin-phonon coupling term:
\begin{align}
F_\mathrm{s-ph} = \int [-\lambda Q\cos(\theta_A-\theta_B) + \frac{\epsilon Q^2}{2}]d^3x.
\end{align}

We define a pair of new variables:
\begin{subequations}
\begin{align}
\psi = \frac{\theta_A + \theta_B}{2};\quad \theta = \theta_A - \theta_B.
\end{align}
The Landau free energy then separate into two independent pieces: 
\begin{align}
F = F_\psi + F_{\theta Q}.
\end{align}
$F_\psi$ is the Landau free energy concerning the spontaneous breaking of the spin rotational symmetry:
\begin{align}
F_\psi = \frac{K_\psi}{2} \int (\nabla\psi)^2 d^3x.
\end{align}
$K_\psi = 2K (1+\kappa)$ is the stiffness of $\psi$. $F_{\theta Q}$ captures the spin-Peierls instability of the system:
\begin{align}
F_{\theta Q} = \int [\frac{K_\theta}{2}(\nabla\theta)^2 - \lambda Q \cos\theta + \frac{\epsilon Q^2}{2}] d^3x,
\end{align}
\end{subequations}
where the stiffness $K_\theta = K(1-\kappa)/2$. 

We discuss briefly on the connection of the effective model parameters to microscopic model parameters. To this end, we consider a microscopic Hamiltonian:
\begin{align}
H = J_1 \sum_{\langle ij\rangle} \mathbf{S}_i\cdot\mathbf{S}_j + J_2 \sum_{\langle\langle ij\rangle\rangle} \mathbf{S}_i\cdot\mathbf{S}_j - J' \sum_{i\in A}\sum_n Q_i \eta_n \mathbf{S}_i\cdot\mathbf{S}_{i+n} + \frac{k}{2}\sum_{i\in A}Q^2_i.
\end{align}
Here, the first and second terms sum over the nearest and second neighbor pairs in the BCC lattice. In the third term, we sum over all bonds $n$ that emanate from an $A$ site, namely $n = (\pm a/2, \pm a/2, \pm a/2)$, where $a$ is the lattice constant. $\eta_n$ is a form factor:
\begin{align}
\eta_n = \left\{\begin{array}{cc}
1 & n=(a/2,a/2,a/2),(a/2,-a/2,-a/2),(-a/2,a/2,-a/2),(-a/2,-a/2,a/2) \\
-1 & n=(-a/2,-a/2,-a/2),(-a/2,a/2,a/2),(a/2,-a/2,a/2),(a/2,a/2,-a/2)  
\end{array}\right. .
\end{align}
Note this model does not break the symmetry between the A and B sub-lattices: the third sum can be rewritten in terms of B sublattices with $Q\to -Q$.

We proceed to coarse-grain the above model by using the gradient expansion. We postulate:
\begin{align}
\mathbf{S}_{i\in A} = \mathbf{N}_A(r_i) e^{iK\cdot r_i};
\quad
\mathbf{S}_{i\in B} = \mathbf{N}_B(r_i) e^{iK\cdot (r_i-n_0)};
\quad
Q_{i} = Q(r_i),
\end{align}
where the right hand sides are smooth functions of the coordinates. $n_0 = (a/2,a/2,a/2)$. $K = (\pi,\pi,\pi)/a$ is the characteristic wave vector  of the N\'{e}el order. To the quadratic order in lattice constant $a$, we have:
\begin{align}
H = \int \frac{d^3x}{a^3} \{\frac{J_2a^2}{2} [(\nabla \mathbf{N}_A)^2 + (\nabla \mathbf{N}_B)^2] - 8J' Q \mathbf{N}_A\cdot\mathbf{N}_B + \frac{k}{2}Q^2\}.
\end{align}
Setting $\mathbf{N}_A = (\cos\theta_A,\sin\theta_A)$ and $\mathbf{N}_B = (\cos\theta_B,\sin\theta_B)$, we obtain the expression of Landau free energy in the main text:
\begin{align}
H = \int d^3x \{\frac{K}{2}[(\nabla\theta_A)^2+(\nabla\theta_B)^2] - \lambda Q\cos(\theta_A-\theta_B) + \frac{\epsilon}{2}Q^2\},
\end{align} 
with
\begin{align}
K = \frac{J_2}{a};\quad \lambda = \frac{8J'}{a^3};\quad \epsilon = \frac{k}{a^3}.
\end{align}
Interestingly, we see that $\kappa = 0$ to the leading order of gradient expansion. It can be generated by integrating out short-ranged fluctuations, or by coarse-graining multi-spin interactions. We also see that $J_1$ term does not contribute at this order (it contributes to the third order in the gradient expansion). 

\section{Additional data for the structure factor}

\begin{figure}
\includegraphics[width = 0.6\columnwidth]{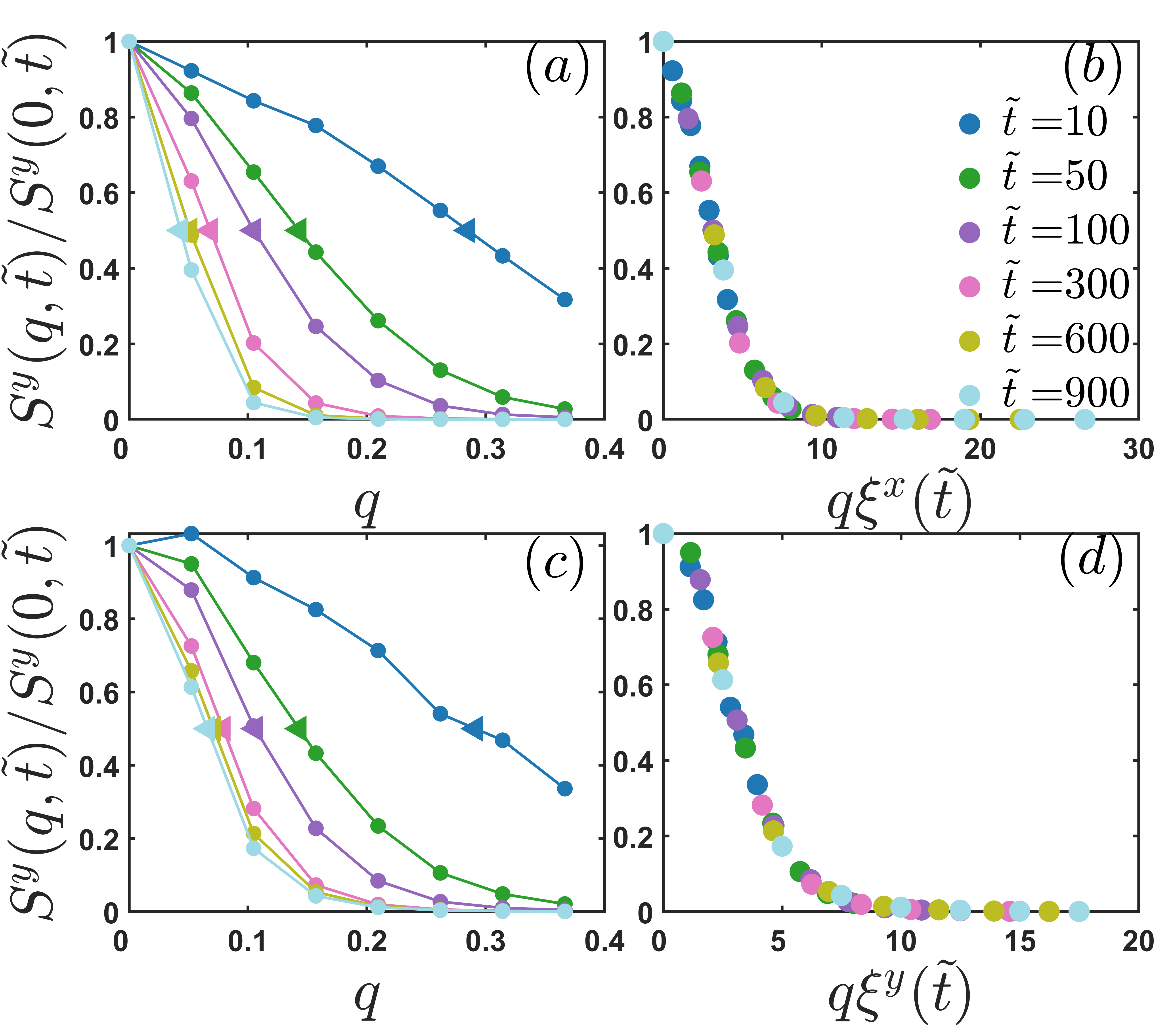}
\caption{(a) The structure factor $S^x(\bm{q})$ along the $(100)$ direction at different times. The phonon relaxation time $\tau = 150$. The system size $L=120$. Triangular symbols mark the half width at half maximum. (b) Collapse of the structure factor at different times. (c)(d) Similar to (a) and (b) but for $S^y(\bm{q})$.}
\label{fig:data_collapse}
\end{figure}

Fig.~\ref{fig:data_collapse}a\&c illustrate the time evolution of the structure factors $S^{x,y}(\bm{q})$ along the $(100)$ direction. We estimate the location of the half width at half maximum, $q_0$, by a simple linear interpolation (triangular markers). Because the Bragg peaks become quite narrow at late time, only two momentum points are useable for the presented system size $L=120$ (data for $\widetilde{t}=600$ and $\widetilde{t}=900$). This results in a larger degrees of uncertainty in determining $q_0$, which, in turn, translates to larger uncertainty in estimating the correlation length $\xi^{x,y}$. 

We may approximately collapse the structure factors at different times by plotting them as functions of $q\xi^{x,y}$ (Fig.~\ref{fig:data_collapse}b\&d). The data collapse of $S^x(\bm{q})$ is consistent with the scaling invariance of the phase ordering process. However, we are unable to make any reliable statements about the Prod's law due to limited system size.

\section{Discretization and non-dimensionalization of the minimal model}

\subsection{Separation of variables}

We begin with the equations of motion from the main text:
\begin{align}
\dot{\theta}_{A,B} = -\Gamma_\mathrm{s} \frac{\delta F}{\delta \theta_{A,B}} + \xi_{A,B};
\quad
\dot{Q} = -\Gamma_\mathrm{ph} \frac{\delta F}{\delta Q} + \xi_Q
\end{align}
Here, $\xi_{A,B,Q}$ are the Gaussian white noise fields. Their second order moments are given by:
\begin{align}
\langle \xi_A (x,t)\xi_A(x',t')\rangle = \langle \xi_B (x,t)\xi_B (x',t')\rangle = 2k_BT \Gamma_\mathrm{s}  \delta^{(3)}(x-x') \delta(t-t');
\nonumber\\
\langle \xi_Q(x,t) \xi_Q(x',t') \rangle = 2k_BT \Gamma_\mathrm{ph} \delta^{(3)}(x-x') \delta(t-t').
\end{align}
Inserting the expression for the Landau free energy into the functional derivative:
\begin{align}
F = \int  \{ \frac{K}{2} [(\nabla\theta_A)^2 + (\nabla\theta_B)^2 + 2\kappa \nabla\theta_A \nabla\theta_B] - \lambda\cos(\theta_A -\theta_B) Q + \frac{\epsilon}{2}Q^2 \} d^3x,
\end{align}
we obtain:
\begin{subequations}
\begin{align}
\dot{\theta}_{A} = -\Gamma_\mathrm{s} (-K\nabla^2\theta_A - \kappa K\nabla^2\theta_B + \lambda \sin(\theta_A-\theta_B)Q) + \xi_A;
\\
\dot{\theta}_{B} = -\Gamma_\mathrm{s} (-K\nabla^2\theta_B - \kappa K\nabla^2\theta_A - \lambda \sin(\theta_A-\theta_B)Q) + \xi_B;
\\
\dot{Q} = -\Gamma_\mathrm{ph} (\epsilon Q - \lambda \cos(\theta_A-\theta_B) ) + \xi_Q.
\end{align}
\end{subequations}

In the next step, we define a pair of new fields:
\begin{align}
\psi = \frac{\theta_A+\theta_B}{2}; 
\quad
\theta = \theta_A - \theta_B.
\end{align}
Summing up the equation of motion for $\theta_A$ and $\theta_B$ yields the equation of motion for $\psi$:
\begin{align}
\dot{\psi} = -\Gamma_s[-K(1+\kappa) \nabla^2\psi]+\xi_\psi  = -\frac{\Gamma_s}{2}(-K_\psi \nabla^2\psi) + \xi_\psi.
\end{align}
Here, $K_\psi = 2K(1+\kappa)$ is the stiffness of $\psi$. $\xi_\psi = (\xi_A+\xi_B)/2$. Its second moment is given by:
\begin{align}
\langle \xi_\psi(x,t) \xi_\psi(x',t')\rangle = 2k_BT \times\frac{\Gamma_s}{2}\delta^{(3)}(x-x') \delta(t-t').
\end{align}

Subtracting off the equation of motion of $\theta_B$ from that of $\theta_A$ yields the equation of motion for $\theta$:
\begin{align}
\dot{\theta} = -\Gamma_\mathrm{s}[-K(1-\kappa)\nabla^2\theta + 2\lambda\sin\theta\, Q] + \xi_\theta = -2\Gamma_\mathrm{s}(-K_\theta \nabla^2\theta + \lambda \sin\theta\,Q) + \xi_\theta.
\end{align}
Here, $K_\theta = K(1-\kappa)/2$ is the stiffness of $\theta$. $\xi_\theta = \xi_A-\xi_B$. Its second moment is given by:
\begin{align}
\langle \xi_\theta(x,t) \xi_\theta(x',t') \rangle = 2k_BT \times 2\Gamma_s \delta^{(3)}(x-x') \delta(t-t').
\end{align}

Finally, the equation of motion for $Q$ is given by:
\begin{align}
\dot{Q} = -\Gamma_\mathrm{ph}(\epsilon Q - \lambda\cos\theta) + \xi_Q.
\end{align}

We note that the equation of motion for $\theta$ and $Q$ can be cast in the following form:
\begin{align}
\dot{\theta} = -2\Gamma_\mathrm{s} \frac{\delta F}{\delta\theta} + \xi_\theta,
\quad
\dot{Q} = -\Gamma_\mathrm{ph}  \frac{\delta F}{\delta Q} + \xi_Q.
\end{align}
Here, $F$ is the Landau free energy:
\begin{align}
F = \int [\frac{K_\theta}{2} (\nabla\theta)^2 - \lambda\cos\theta \,Q + \frac{\epsilon}{2}Q^2]d^3x.
\end{align}
The variable $\psi$ is dropped from the minimal model. The above set of equations constitute the continuum description of the minimal model used in the main text. In the next subsection, we discretize this model to facilitate numerical simulation. 

\subsection{Discretization}

We discretize the continuous space by a simple cubic grid with lattice spacing $a$. We define:
\begin{align}
\theta_i = \theta(x_i);
\quad
Q_i = Q(x_i)
\end{align}
where $x_i$ are the coordinates of the lattice site $i$. Furthermore, for $x$ inside a cube of size $a$, centered at $x_i$, we have:
\begin{align}
\theta(x) \approx \psi_i;
\quad
Q(x) \approx Q_i.
\end{align}
We first discretize the the Landau free energy:
\begin{align}
F  &= \int [\frac{K_\theta}{2a^2} (a\nabla\theta)^2 - \lambda\cos\theta Q + \frac{\epsilon}{2}Q^2] a^3 \frac{d^3x}{a^3}
\nonumber\\
&\approx -\sum_{\langle ij\rangle} (K_\theta a)\cos(\theta_i - \theta_j) + \sum_i a^3(-\lambda \cos\theta_i Q_i + \frac{\epsilon}{2}Q^2_i).
\end{align}
The equation of motion for $\theta_i$ is obtained by averaging over the small cube centered at $x_i$:
\begin{align}
\dot{\theta}_i = -2\Gamma_s \frac{1}{a^3} \int_i \frac{\delta F}{\delta \theta(x)} d^3x + \frac{1}{a^3}\int_i \xi_\theta(x,t) d^3x.
\end{align}
The integral is over the said cube. The first integral on the right hand side represents the change of $F$ due to a uniform increase of $\theta(x)$ inside the cube: $\theta(x) \to \theta(x) + \delta \theta_i$. Therefore:
\begin{align}
\int_i \frac{\delta F}{\delta \theta(x)} d^3x \approx \frac{\partial F}{\partial \theta_i}.
\end{align}
Define:
\begin{align}
\xi_{\theta,i} = \frac{1}{a^3}\int_i \xi_\theta (x,t) d^3x.
\end{align}
The second moment of $\xi_{\theta,i}$ is given by:
\begin{align}
\langle \xi_{\theta,i} (t) \xi_{\theta,j} (t')\rangle = 2k_BT \times \frac{2\Gamma_\mathrm{s}}{a^6} \delta(t-t') \int_i d^3x \int_j d^3x' \delta^{(3)}(x-x') = 2k_BT \times \frac{2\Gamma_\mathrm{s}}{a^3}\delta_{ij} \delta(t-t').
\end{align}

We thus have obtained the following discretized Langevin equation:
\begin{align}
\dot{\theta}_i = -\frac{2\Gamma_s}{a^3} \frac{\partial F}{\partial \theta_i} + \xi_{\theta,i},
\quad
\langle \xi_{\theta,i} (t) \xi_{\theta,j} (t' ) \rangle = 2k_BT\times \frac{2\Gamma_\mathrm{s}}{a^3} \delta_{ij} \delta(t-t').
\end{align}
The discretized form of the equations of motion for $Q$ is obtained in the same vein:
\begin{align}
\dot{Q}_i = -\frac{\Gamma_\mathrm{ph}}{a^3}\frac{\partial F}{\partial Q_i} + \xi_{Q,i};
\quad
\langle \xi_{Q,i} \xi_{Q,j}\rangle = 2k_BT\times \frac{\Gamma_\mathrm{ph}}{a^3} \delta_{ij}\delta(t-t').
\end{align}

\subsection{Non-dimensionalization}

In this subsection, we rescale the space and time variables such that the minimal model parameters are dimensionless. We define an energy scale:
\begin{align}
J_\theta = K_\theta a.
\end{align}
Microscopically, this scale corresponds to the spin exchange energy scale. From $J_\theta$, we may define a characteristic displacement $Q_0$:
\begin{align}
a^3\epsilon Q^2_0 = J_\theta \Rightarrow Q_0 = \sqrt{\frac{K_\theta}{\epsilon a^2}}.
\end{align}
Using these, the dimensionless Landau free energy $\widetilde{F}$ is given by:
\begin{align}
\widetilde{F} = \frac{F}{J_\theta} = -\sum_{\langle ij\rangle}\cos(\theta_i - \theta_j) + \sum_i (-g\cos\theta_i \widetilde{Q}_i + \frac{\widetilde{Q}^2_i}{2}).
\end{align}
Here, we have defined the dimensionless phonon coordinate: $\widetilde{Q}_i = Q_i / Q_0$. $g$ is the dimensionless coupling constant: $g = \lambda / \epsilon Q_0 = \lambda \sqrt{a^2 / (\epsilon K_\theta)}$. 

We turn to the equation of motion for $\theta$.  We define a microscopic time scale for $\theta$ by:
\begin{align}
\tau^{-1}_\theta = \frac{2\Gamma_\mathrm{s}}{a^3} J_\theta = \frac{2\Gamma_\mathrm{s}K_\theta}{a^2}.
\end{align}
The dimensionless time variable $\widetilde{t}$ is measured in unit of $\tau_\theta$:
\begin{align}
\widetilde{t} = \frac{t}{\tau_\theta}.
\end{align}
The equation of motion for $\theta$ now becomes:
\begin{align}
\frac{1}{\tau_\theta}\frac{d\theta_i}{d\widetilde{t}} = - \frac{1}{\tau_\theta}\frac{\partial \widetilde{F}}{\partial \theta_i} + \xi_{\theta,i} \Rightarrow \frac{d\theta_i}{d\widetilde{t}} = -\frac{\partial \widetilde{F}}{\partial \theta_i} + \nu_{i},
\end{align}
where the dimensionless white noise,
\begin{align}
\nu_i(\widetilde{t}) = \tau_\theta \xi_{\theta,i}(\widetilde{t}\tau_\theta).
\end{align}
Its second moment reads:
\begin{align}
\langle \nu_i(\widetilde{t}) \nu_j(\widetilde{t}')\rangle = \tau^2_\theta\times 2k_BT \times \frac{2\Gamma_s}{a^3} \delta_{ij}\delta(\widetilde{t}\tau_\theta-\widetilde{t}'\tau_\theta) = 2\widetilde{T} \delta_{ij} \delta(\widetilde{t}-\widetilde{t}'),
\end{align}
where 
\begin{align}
\widetilde{T} = \frac{k_BT}{J_\theta} = \frac{k_BT}{K_\theta a},
\end{align}
is the dimensionless temperature.

The equation of motion for $\widetilde{Q}_i$ is given by:
\begin{align}
\frac{Q_0}{\tau_\theta}\frac{d\widetilde{Q}_i}{d\widetilde{t}} = -\frac{\Gamma_\mathrm{ph} J_\theta}{a^3 Q_0}\frac{\partial \widetilde{F}}{\partial \widetilde{Q}_i} + \xi_{Q,i}.\Rightarrow
\frac{d\widetilde{Q}_i}{d\widetilde{t}} = -\frac{1}{\tau} \frac{\partial \widetilde{F}}{\partial \widetilde{Q}_i} + \eta_i.
\end{align}
Here, $1/\tau$ is a dimensionless ratio of relaxation rates: 
\begin{align}
\frac{1}{\tau} = \frac{\tau_\theta \Gamma_\mathrm{ph} J_\theta}{a^3 Q^2_0} = \tau_\theta \Gamma_\mathrm{ph}\epsilon = \frac{a^2\Gamma_\mathrm{ph}\epsilon}{2\Gamma_\mathrm{s}K_\theta}.
\end{align}
We have also defined a dimensionless white noise:
\begin{align}
\eta_i(\widetilde{t}) = \frac{\tau_\theta}{Q_0} \xi_{Q,i}(\widetilde{t}\tau_\theta).
\end{align}
Its second moment is given by:
\begin{align}
\langle \eta_i (\widetilde{t}) \eta_j(\widetilde{t}') \rangle = \frac{\tau^2_\theta}{Q^2_0}\times 2k_BT\times \frac{\Gamma_\mathrm{ph}}{a^3}\delta_{ij}\delta(\widetilde{t}\tau_\theta-\widetilde{t}'\tau_\theta) = \frac{2\widetilde{T}}{\tau}\delta_{ij}\delta(\widetilde{t}-\widetilde{t}').
\end{align}

To summarize, the non-dimensionalized equations of motion read:
\begin{subequations}
\begin{align}
\frac{d\theta_i}{d\widetilde{t}} = -\frac{\partial \widetilde{F} }{\partial \theta_i} + \nu_{i};
\quad
\frac{d\widetilde{Q}_i}{d\widetilde{t}} = -\frac{1}{\tau}\frac{\partial \widetilde{F} }{\partial \widetilde{Q}_i} + \eta_{i}.
\end{align}
The dimensionless Landau free energy is given by:
\begin{align}
\widetilde{F} = -\sum_{\langle ij\rangle}\cos(\theta_i - \theta_j) + \sum_i -g\cos\theta_i \widetilde{Q}_i + \frac{\widetilde{Q}^2_i}{2}.
\end{align}
The second moments of the white noises are given by:
\begin{align}
\langle \nu_i(\widetilde{t}) \nu_j(\widetilde{t}')\rangle = 2\widetilde{T}\delta_{ij}\delta(\widetilde{t} - \widetilde{t}').
\quad
\langle \eta_i (\widetilde{t}) \eta_j(\widetilde{t}')\rangle = \frac{2\widetilde{T}}{\tau}\delta_{ij}\delta(\widetilde{t} - \widetilde{t}').
\end{align}
\end{subequations}
The above are the equations used in the numerical simulation.

\section{Representative spin configuration snapshots}

\begin{figure}
\includegraphics[width = 0.8\columnwidth]{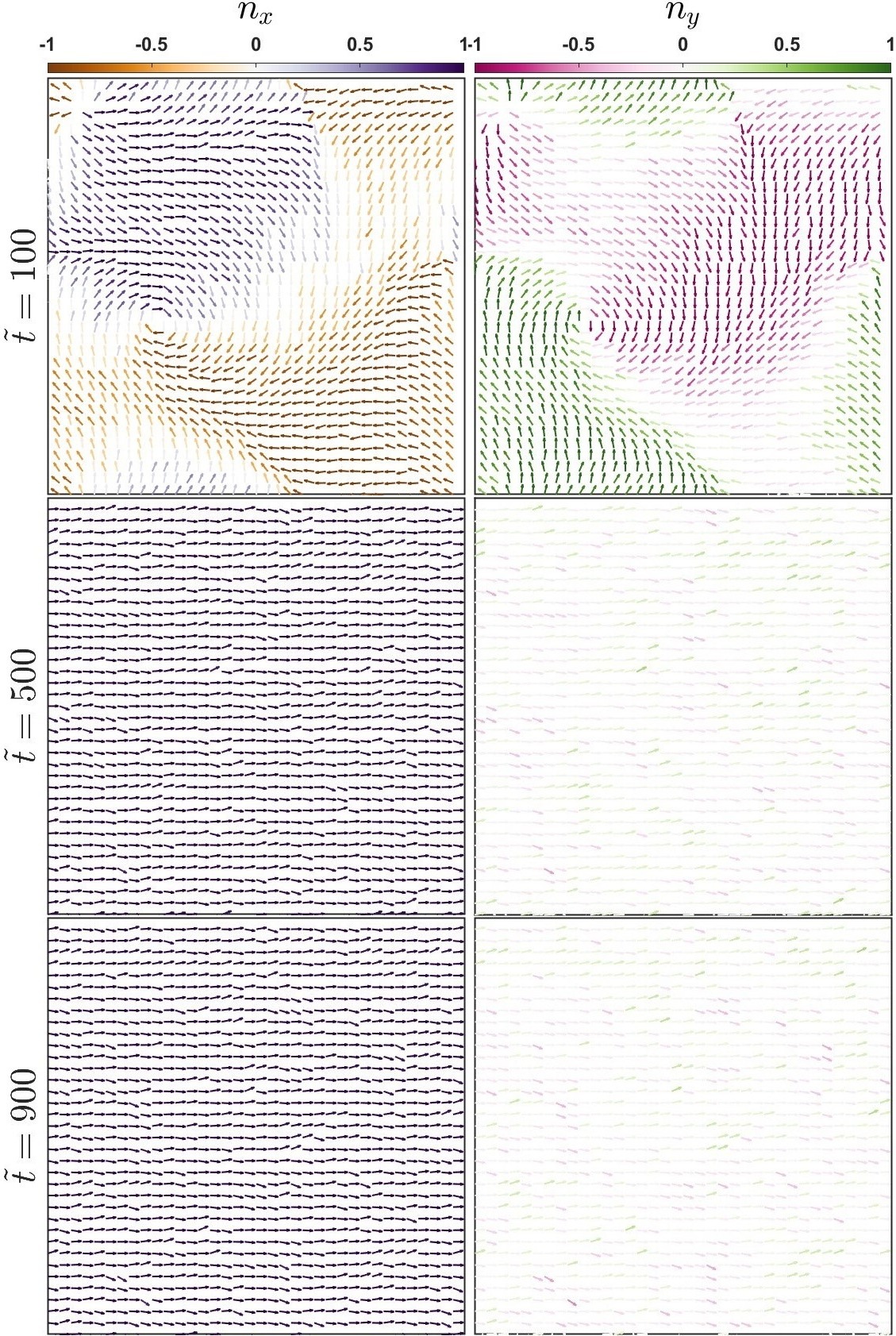}\
\caption{Snapshots of representative spin configurations at time $\widetilde{t} = 100$, $500$, and $900$. The model parameters $g = 0.2$, $\tau = 150$, $\widetilde{T} = 0.1$. The initial conditions are identical to the one used in the main text. The system size $L=36$. The left column highlights the $x$ component $n_x = \cos\theta$. whereas the right column highlights the $y$ component $n_y = \sin\theta$ of the same data.}
\label{fig:snapshot_L36}
\end{figure}

\begin{figure}
\includegraphics[width = 0.8\columnwidth]{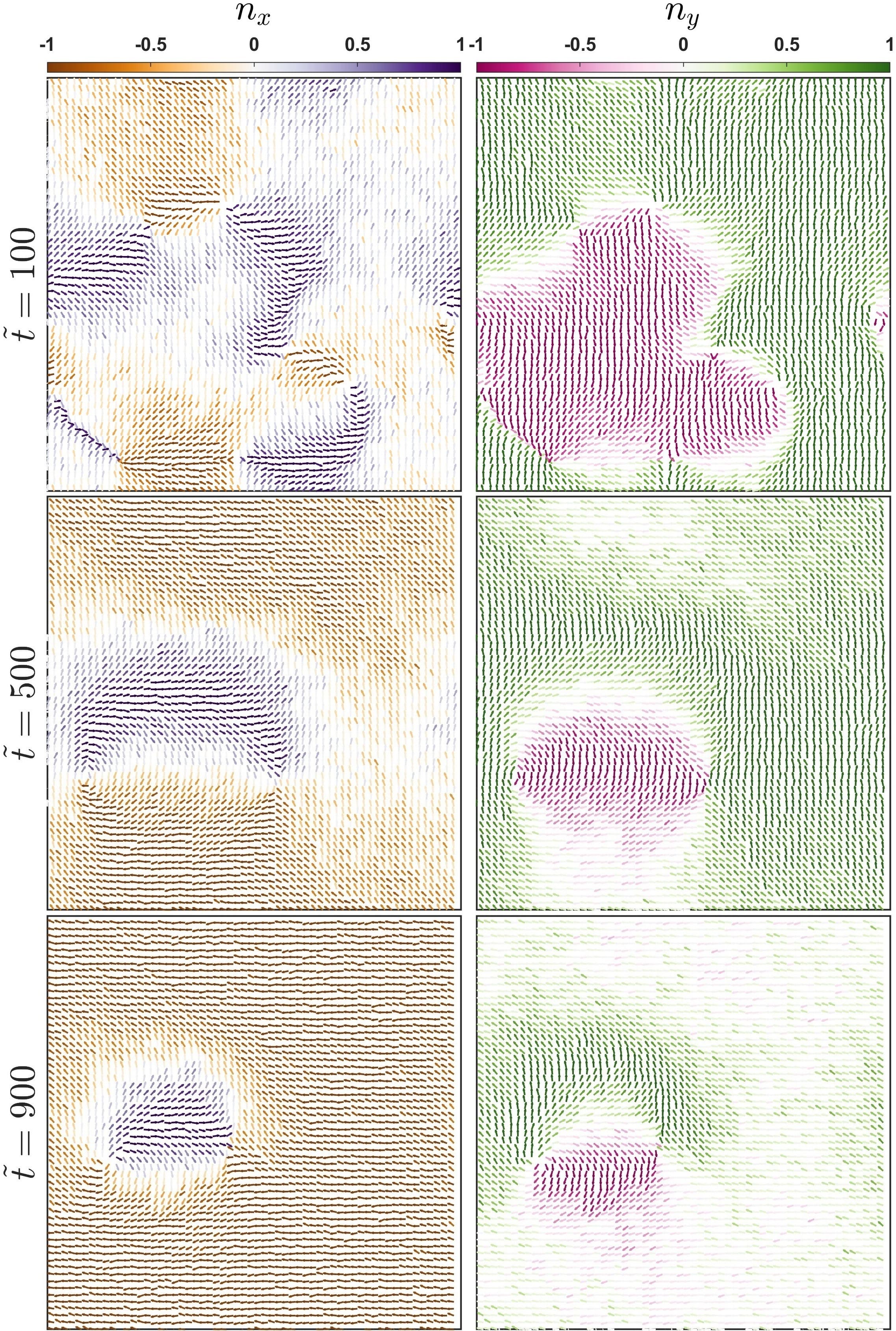}
\caption{Same as Fig.~\ref{fig:snapshot_L36} but for larger system size $L=60$.}
\label{fig:snapshot_L60}
\end{figure}

Fig.~\ref{fig:snapshot_L36} shows the representative spin configurations in a run with small system size $L=36$, taken from an arbitrarily chosen cross section of the whole three dimensional lattice. We see that the early time $\widetilde{t} = 100$ snapshot shows the coexistence of both collinear (spin $\parallel x$) and non-collinear (spin $\parallel y$) states. At late time, the system is taken over by a single domain of the collinear state with $n_x\approx 1$. 

For larger system size ($L=60$), we observe the system has both $n_x \approx 1$ and $n_x\approx -1$ domains at late time. The non-collinear states exist near the domain wall. Interestingly, within the $n_x$ domain wall where $n_y$ dominates, there are also domain structure with $n_y \approx 1$ and $n_y \approx -1$. These ``domain walls within a domain wall" in fact are the vortex strings. The slow motion of the defects leads to the slow decrease in the structure factor $S^y(q=0)$.

\section{Impact of initial correlation on dynamics \label{sec:initial_corr}}

\begin{figure}
\includegraphics[width = \columnwidth]{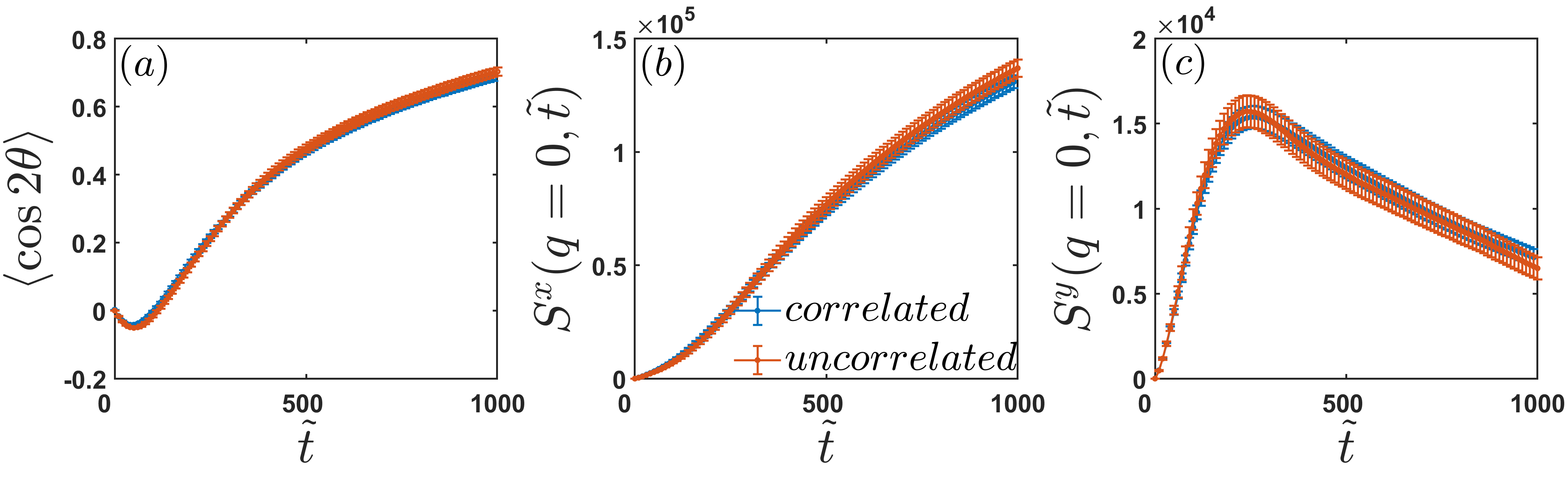}
\caption{Comparison of numerical results from the uncorrelated initial states (red) and correlated initial states (blue). (a) The local anisotropy. (b) Structure factor $S^x(q=0)$ as a function of time $\widetilde{t}$. (c) Structure factor $S^y(q=0)$ as a function of time $\widetilde{t}$.}
\label{fig:compare_initial_state}
\end{figure}

In the main text, we set the initial dimensionless temperature $\widetilde{T}_i = 3$. As this temperature is much higher than the energy scale of spin exchange interaction ($1$) and the energy scale of the spin-phonon coupling ($g=0.2$), we omit the correlations between the spins, as well as the correlation between the spins and the phonons. This resulted in an uncorrelated initial state where the $\theta_i$ are drawn independently from a uniform distribution and the $\widetilde{Q}_i$ are drawn from a Gaussian distribution whose variance is fixed by the equipartition theorem, $\langle \widetilde{Q}^2_i\rangle = \widetilde{T}_i = 3$. 

Our choices of this approximate initial condition is to reduce the numerical costs. In this section, we present the results from a numerical simulation with the initial condition drawn from the Gibbs ensemble with $\widetilde{T}_i = 3$. All other choices of parameters are identical to that of the main text. The results are shown in Fig.~\ref{fig:compare_initial_state}. We find that, for all key observables (the local anisotropy and the structure factors), the results are quantitatively close to each other. This justifies the approximation used in the main text.

\section{Order by quantum and thermal disorder effect}

The quantum/thermal order by disorder effect favors the collinear magnetic order. Therefore, they would compete with the non-equilibrium phonons. Their impact can be captured by adding a term of the form:
\begin{align*}
-\Delta \cos^2 (\theta_A-\theta_B)
\end{align*} 
to our model. Here, $\Delta>0$ reflects the overall energy scale of the selection effect due to quantum/thermal fluctuations. Its value depends on the microscopic details of the model. In this section, we compute $\Delta$ and relate it to the microscopic model parameters. 

We consider a quantum $J_1$-$J_2$ XY model on the BCC lattice. Its Hamiltonian is given by:
\begin{align}
H = \frac{1}{2I} \sum_{i}\sum_{\alpha=A,B}  L^2_{i\alpha}+ J_2 \sum_{\langle ij\rangle}\sum_{\alpha=A,B}\cos(\theta_{i\alpha}-\theta_{j\alpha}) + J_1\sum_{i\in A}\sum_{n} \cos(\theta_{iA}-\theta_{i+n,B}).
\end{align}
Here, $\alpha = A,B$ labels the two simple cubic sublattices. $\theta_{i\alpha}$ and $L_{i\alpha}$ are respectively the angle and the angular momentum of the planar rotor on site $i$ of the sublattice $\alpha$.  $I>0$ is the rotational inertia. The second term sums over all the nearest neighboring pairs within the same sublattice, where as the third term sums over all the neighboring pairs between the sublattice A and B. $n$ are the vectors that point from an A site to the neighboring B sites. Setting the lattice parameter to $1$, $n$ are given by $n = (\pm \frac{1}{2}, \pm\frac{1}{2}, \pm\frac{1}{2})$. We use the coordinate system whose origin is at an A site. 

The classical ground states are two independent N\'{e}el orders on the two sublattices. We may parametrize the N\'{e}el order on the sublattice A by an angle $\theta_A$, which is defined to be the angle of the site at the origin; likewise, we parametrize the N\'{e}el order on sublattice B by $\theta_B$, which is defined to be the angle of the site at $(1/2,1/2,1/2)$. The classical ground state energy ($I\to \infty$ limit) is independent of $\theta_A$ and $\theta_B$, indicating the accidental degeneracy of the classical ground state.

We now compute the correction to the ground state energy due to zero point motions of the spin wave modes. To this end, we write $\theta_{i\alpha} = \theta^{(0)}_{i\alpha} + \delta \theta_{i\alpha}$, where $\theta^{(0)}_{i\alpha}$ is the ground state value, and $\delta\theta_{i\alpha}\ll 1$ is the small deviation about the ground state. Expanding the Hamiltonian to quadratic order in $\delta\theta_{i\alpha}$, we obtain:
\begin{align}
H = \frac{1}{2I}\sum_{i\alpha}L^2_{i\alpha} + \frac{J_2}{2}\sum_{\langle ij\rangle\alpha}(\delta\theta_{i\alpha}-\delta\theta_{j\alpha})^2 - \frac{J_1}{2}\cos(\theta_A-\theta_B)\sum_{i\in A}\sum_n \eta_n (\delta\theta_{iA} - \delta\theta_{i+n,B})^2.
\end{align}
$\eta_n$ is the former factor defined in Section I. The above Hamiltonian is diagonal in the momentum basis:
\begin{align}
H = \sum_q \frac{1}{2I}L_{-q\alpha}L_{q\alpha}  + \frac{1}{2}
\begin{pmatrix}
\theta_{-qA} & \theta_{-qB}
\end{pmatrix}
\begin{pmatrix}
J_2\epsilon_q & J_1\cos(\theta_A-\theta_B)\eta_q \\
J_1\cos(\theta_A-\theta_B)\eta^\ast_q & J_2\epsilon_q
\end{pmatrix}
\begin{pmatrix}
\theta_{qA} \\
\theta_{qB}
\end{pmatrix}.
\end{align}
Here,
\begin{align}
\epsilon_q = 4(\sin^2\frac{q_x}{2}+\sin^2\frac{q_y}{2}+\sin^2\frac{q_z}{2});
\quad
\eta_q = -8i\sin\frac{q_x}{2}\sin\frac{q_y}{2}\sin\frac{q_z}{2}.
\end{align}
Diagonalizing the above Hamiltonian, we find two branches of spin waves; their dispersion relation are given by:
\begin{align}
\omega_{q\pm} = \sqrt{\frac{K_{q\pm}}{I}},\quad K_{q\pm} = J_2\epsilon_q \pm J_1\cos(\theta_A-\theta_B)|\eta_q|.
\end{align}
from which we obtain the zero point energy density:
\begin{align}
\frac{E}{V} = \frac{\hbar}{2}\int\frac{d^3q}{(2\pi)^3}(\omega_{q+}+\omega_{q-}),
\end{align}
where the domain of integration is the first Brillouin zone. To make analytic progress, we perform a Taylor expansion on $J_1/J_2$:
\begin{align}
\frac{E}{V} = \hbar\sqrt{\frac{J_2}{I}} \int\frac{d^3q}{(2\pi)^3} \epsilon^{1/2}_q - \frac{\hbar}{8}\sqrt{\frac{J_2}{I}} \frac{J^2_1}{J^2_2}\cos^2(\theta_A-\theta_B)\int\frac{d^3q}{(2\pi)^3} \frac{|\eta_q|^2}{\epsilon^{3/2}_q}.
\end{align}
We may read off $\Delta$ from the ground state energy:
\begin{align}
\Delta_\mathrm{obqd} = 0.0407\times\hbar\sqrt{\frac{J_2}{I}}(\frac{J_1}{J_2})^2.
\end{align}

Similar calculations can be done for the order by thermal disorder. In this case, the relevant quantity is the free energy contribution from the thermal fluctuations in the spin wave:
\begin{align}
F = \frac{k_BT}{2}\int \frac{d^3q}{(2\pi)^3}(\ln K_{q+} + \ln K_{q-}).
\end{align}
Expanding to the leading order in $J_1/J_2$, we obtain:
\begin{align}
F = k_BT \int \frac{d^3q}{(2\pi)^3}\ln\epsilon_q - \frac{k_BT}{2}\frac{J^2_1}{J^2_2}\cos^2(\theta_A-\theta_B)  \int \frac{d^3q}{(2\pi)^3} \frac{|\eta_q|^2}{\epsilon^2_q},
\end{align}
from which we extract $\Delta$:
\begin{align}
\Delta_\mathrm{obtd} = 0.0581\times k_BT(\frac{J_1}{J_2})^2.
\end{align}

\section{Numerical simulation of the two order parameter model \label{sec:two_rotor}}

\subsection{Discretization and non-dimensionalization}

The minimal model discussed in the main text omits the N\'{e}el order parameter $\psi$, which reflects the spontaneous symmetry breaking of the true spin $U(1)$ symmetry. As a result, this minimal model captures the selection of competing magnetic orders after the quench but cannot describe the accompanying revival of the magnetic order. In this section, we present the numerical simulation of the model with both $\theta_A$ and $\theta_B$ degrees of freedom (Eqs.~(1) and (2) in the main text), which treat these two non-equilibrium processes on equal footing. 

The equations of motion in the continuous space are given by:
\begin{subequations}
\begin{align}
\dot{\theta}_{A,B} = -\Gamma_\mathrm{s} \frac{\delta F}{\delta \theta_{A,B}} + \xi_{A,B};
\quad
\dot{Q} = -\Gamma_\mathrm{ph} \frac{\delta F}{\delta Q} + \xi_Q
\end{align}
Here, $\xi_{A,B,Q}$ are the Gaussian white noise fields. Their second order moments are given by:
\begin{align}
\langle \xi_A (x,t)\xi_A(x',t')\rangle = \langle \xi_B (x,t)\xi_B (x',t')\rangle = 2k_BT \Gamma_\mathrm{s}  \delta^{(3)}(x-x') \delta(t-t');
\nonumber\\
\langle \xi_Q(x,t) \xi_Q(x',t') \rangle = 2k_BT \Gamma_\mathrm{ph} \delta^{(3)}(x-x') \delta(t-t').
\end{align}
The Landau free energy is given by:
\begin{align}
F = \int  \{ \frac{K}{2} [(\nabla\theta_A)^2 + (\nabla\theta_B)^2] - \lambda\cos(\theta_A -\theta_B) Q + \frac{\epsilon}{2}Q^2 \} d^3x,
\end{align}
\end{subequations}
Here, we set $\kappa = 0$ for simplicity.

For numerical purposes, we apply the discretization and nondimensionalization process to the above model. Here, we simply quote the result. We define an energy scale:
\begin{align}
J = Ka,
\end{align}
as well as the spin time scale
\begin{align}
\tau^{-1}_\mathrm{s} = \frac{\Gamma_\mathrm{s}K}{a^2}.
\end{align}
Here, $a$ is the short-distance cutoff. The discretized, dimensionless free energy reads:
\begin{align}
\widetilde{F} = -\sum_{\langle ij\rangle} \cos(\theta_{iA}-\theta_{jA}) + \cos(\theta_{iB}-\theta_{jB}) + \sum_i -g \cos(\theta_{iA}-\theta_{iB}) \widetilde{Q}_i + \frac{\widetilde{Q}^2_i}{2}.
\end{align}
Here, $\widetilde{Q}_i = Q/Q_0$ is the dimensionless phonon displacement; $Q_0 = \sqrt{J/(\epsilon a^3)} = \sqrt{K/(\epsilon a^2)}$ is a characteristic phonon displacement scale. $g = \lambda/(\epsilon Q_0) = \lambda\sqrt{a^2/K\epsilon}$ is the dimensionless coupling.

We rescale time variable as:
\begin{align}
\widetilde{t} = \frac{t}{\tau_\mathrm{s}}.
\end{align}
The equations of motion read:
\begin{align}
\frac{d\theta_{iA}}{d\widetilde{t}} = -\frac{\partial\widetilde{F}}{\partial\theta_{iA}} + \nu_{iA};
\quad
\frac{d\theta_{iB}}{d\widetilde{t}} = -\frac{\partial\widetilde{F}}{\partial\theta_{iB}} + \nu_{iB};
\quad
\frac{d\widetilde{Q}_{i}}{d\widetilde{t}} = -\frac{1}{\tau}\frac{\partial\widetilde{F}}{\partial \widetilde{Q}_{i}} + \eta_{i};
\end{align}
Here, $\tau$ is ratio of relaxation time scales:
\begin{align}
\frac{1}{\tau} = \epsilon \Gamma_\mathrm{ph} \tau_\mathrm{s} = \frac{a^2\epsilon \Gamma_\mathrm{ph}}{K\Gamma_\mathrm{s}}.
\end{align}
The second moments of the white noises are given by:
\begin{align}
\langle \nu_{Ai}(\widetilde{t}) \nu_{Aj}(\widetilde{t}') \rangle = \langle \nu_{Bi}(\widetilde{t}) \nu_{Bj}(\widetilde{t}') \rangle = 2\widetilde{T} \delta_{ij} \delta(\widetilde{t}-\widetilde{t}');
\quad
\langle \eta_{i}(\widetilde{t}) \eta_{j}(\widetilde{t}') \rangle = \frac{2\widetilde{T}}{\tau} \delta_{ij} \delta(\widetilde{t}-\widetilde{t}').
\end{align}

\subsection{Numerical results}

\begin{figure}
\includegraphics[width = 0.6\columnwidth]{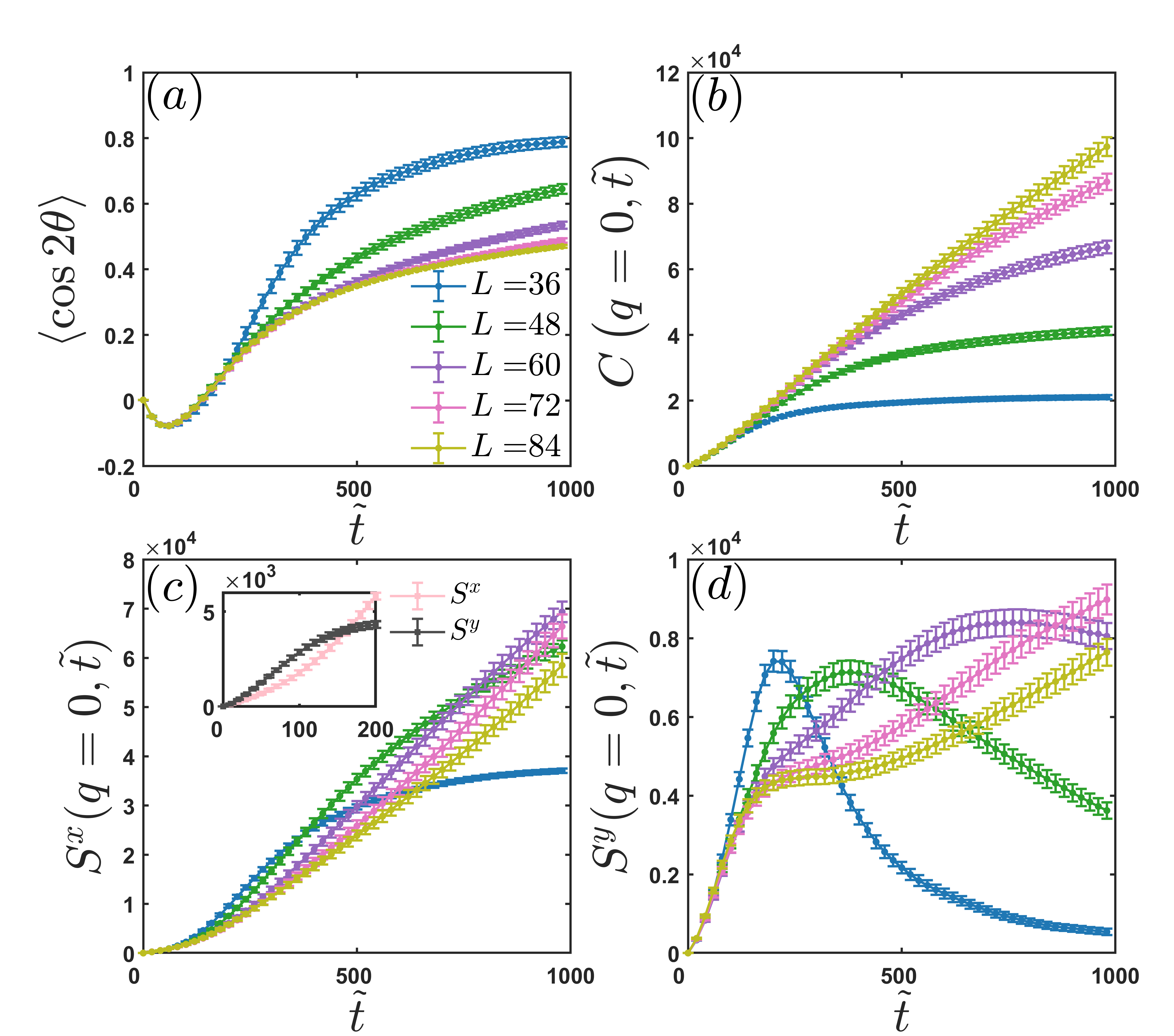}
\caption{Numerical simulation of the model with both $\theta_A$ and $\theta_B$ degrees of freedom. (a) The local spin anisotropy as a function of dimensionless time $\widetilde{t}$. (b) The structure factor $C(q=0)$, which measures the development of the N\'{e}el order, as a function of $\widetilde{t}$. (c) The structure factor $S^x(q=0)$. (d) The structure factor $S^y(q=0)$. The inset shows a comparison of $S^x$ and $S^y$ at early times.}
\label{fig:two_rotor_model}
\end{figure}

In this subsection, we present the numerical simulation of the two order parameter model (Fig.~\ref{fig:two_rotor_model}). We set the dimensionless coupling $g=0.2$, and the ratio of relaxation times $\tau = 150$. We use the same uncorrelated initial state as the main text, which has already been justified by Sec.~\ref{sec:initial_corr}. The spins $\theta_A$ and $\theta_B$ are drawn independently from a uniform distribution on $[0,2\pi)$; the phonons $\widetilde{Q}$ are drawn from the Gaussian distribution with variance $\widetilde{T}_i = 3$.

The definition of local anisotropy $\langle \cos2\theta\rangle$, and the structure factors $S^{x,y}(q)$ are identical to that of the main text. As these observables are silent about the development of the N\'{e}el order (the $\psi$ in the main text), we define the following structure factor:
\begin{align}
C(\mathbf{q},\widetilde{t}) = \frac{1}{L^3}\sum_{i,j} [\langle n_{iA}(\widetilde{t}) \cdot n_{jA}(\widetilde{t}) \rangle +  \langle n_{iB}(\widetilde{t}) \cdot n_{jB}(\widetilde{t}) \rangle ] e^{-i\mathbf{q}\cdot (\mathbf{r}_i-\mathbf{r}_j)}.
\end{align}
Here, $n_{iA} = (\cos\theta_{iA},\sin\theta_{iA})$ and  $n_{iB} = (\cos\theta_{iB},\sin\theta_{iB})$. This observable probes the spin correlation within each sublattice. In particular, $C(q=0)$ is the height of the magnetic Bragg peak. 

We see that the structure factor $C(q=0)$ grows in time,  reflecting the development of the N\'{e}el order (Fig.~\ref{fig:two_rotor_model}b). Crucially, the local anisotropy (Fig.~\ref{fig:two_rotor_model}a) and the structure factors $S^{x,y}(q=0)$ (Fig.~\ref{fig:two_rotor_model}c\& d) exhibit qualitatively similar behaviors as that of the minimal model. This justifies the minimal model used in the main text. Interestingly, $S^y(q=0)$ for large system size ($L=84$) shows a kink near the spin reorientation and continues growing afterwards. By contrast, for the minimal model, it saturates after the reorientation. We heuristically understand the continued growth of $S^y(q=0)$ in the two order parameter model as the result of the N\'{e}el order growth, which enhanced the remnant non-collinear short-range order.

\section{Numerical simulation of the Landau-Ginzburg model}

The model considered in the main text has made the simplifying assumption that the amplitude of the magnetic moment is fixed. This is motivated by the observation that the the selection of magnetic order is intimated connected to the transverse motion of the order parameter. By contrast, in the phenomenological study of the transient photon-exicted dynamics of long-range ordered systems, it is more common to use the Landau-Ginzburg model, where the amplitude of the order parameter is allowed to vary. In this section, we construct and simulate a Landau-Ginzburg model, which is an extension of the model used in the main text.

\subsection{Construction of the model}

We describe the N\'{e}el order on the A and B sublattices by two complex numbers, $\psi_A$ and $\psi_B$. The modulus of $\psi_{A,B}$ represent the magnitude of the magnetic moment, whereas the phase angles parametrize the orientation of the N\'{e}el order. We describe the phonon displacement by real scalar $Q$, similar to the model considered in the main text.

The symmetry transformation properties of $\psi_A$ and $\psi_B$ are as follows. Under spin rotation by $\alpha$, $\psi_{A,B} \to \psi_{A,B}\exp(i\alpha)$. Under time reversal, $\psi_{A,B} \to -\psi_{A,B}$. Under spatial inversion with respect to an A site, $\psi_A\to \psi_A$ but $\psi_B \to -\psi_B$. Finally, half translation along $[111]$ induces $\psi_B \to \psi_A$ and $\psi_A\to -\psi_B$. These symmetries are sufficient to constrain the form of the Landau free energy.

The Landau free energy density consists of two pieces:
\begin{subequations}
\begin{align}
f = f_\mathrm{s} + f_\mathrm{ph}.
\end{align}
The spin part reads:
\begin{align}
f_\mathrm{s} = \frac{K}{2}(|\nabla \psi_A|^2 + |\nabla \psi_B|^2) + \frac{A}{2} (|\psi_A|^2+|\psi_B|^2) + \frac{B}{4}(|\psi_A|^2+|\psi_B|^2)^2 - C |\psi_A|^2|\psi_B|^2.
\end{align}
The parameter $K>0$ controls the stiffness of the order parameters. $A$ controls the thermodynamic phase of the theory; in the mean field limit, the system is in the ordered phase when $A<0$, and disordered phase when $A>0$. $B>0$ ensures the stability of the theory. $C$ represents the coupling between the magnitude of the N\'{e}el orders; we set $C>0$, reflecting the fact that the two N\'{e}el orders tend to coexist. We note that the quartic term of the form $(\psi^\ast_A)^2 \psi^2_B + (\psi^\ast_B)^2 \psi^2_A$ is also allowed by symmetry; however, this term can be generated by the phonons and thus omitted in $f_\mathrm{s}$.

The phonon part reads:
\begin{align}
f_\mathrm{ph} = -\lambda Q (\psi^\ast_A \psi^{\phantom\ast}_B+\psi^\ast_B \psi^{\phantom\ast}_A) + \frac{\epsilon}{2}Q^2.
\end{align}
Here, $\epsilon>0$ is the elastic constant of the phonon. $\lambda$ is the coupling between phonon and the magnetic order parameter. We set $\lambda>0$ without loss of generality. The total free energy $F = \int f d^3x$. 
\end{subequations}

\begin{subequations}
The equations of motion for $\psi_{A,B}$ read:
\begin{align}
\dot{\psi}_A = -\Gamma_\mathrm{s}\frac{\delta F}{\delta \psi^\ast_A} + \xi_{A};
\quad
\dot{\psi}_B = -\Gamma_\mathrm{s}\frac{\delta F}{\delta \psi^\ast_B} + \xi_{B}.
\end{align}
Here, $\xi_{A,B}$ are complex Gaussian white noise fields with the following second moments:
\begin{align}
\langle \xi^\ast_A(x,t) \xi_A(x',t') \rangle = \langle \xi^\ast_B (x,t) \xi_B (x',t') \rangle = 2k_BT \Gamma_\mathrm{s}\delta^{(3)}(x-x') \delta(t-t').
\end{align}
The equation of motion for $Q$ reads:
\begin{align}
\dot{Q} = -\Gamma_\mathrm{ph}\frac{\delta F}{\delta Q} + \xi_Q.
\end{align}
$\xi_Q$ is a real Gaussian white noise field with the second moment:
\begin{align}
\langle \xi_Q(x,t) \xi_Q(x',t')\rangle = 2k_BT\Gamma_\mathrm{ph} \delta^{(3)}(x-x') \delta(t-t').
\end{align}
\end{subequations}

We note that, if we impose the constraint that $\psi_{A,B} = \psi_0 \exp(i\theta_{A,B})$, where $\psi_0$ is the fixed amplitude and $\theta_{A,B}$ the phase angles, the free energy is reduced to the one considered in the main text, with the identification: $K\psi^2_0 \to K$, $2\lambda\psi^2_0 \to \lambda$.

\subsection{Magnetic order selection}

The above model captures the spin-Peierls instability in the equilibrium magnetically ordered phase. To see this, we seek the minimum of $f$ in the mean field limit. We may set $\psi_{A,B}$ and $Q$ to be uniform in space. Minimizing $f$ with respect to $Q$ yields:
\begin{align}
f = \frac{A}{2} (|\psi_A|^2+|\psi_B|^2) + \frac{B}{4}(|\psi_A|^2+|\psi_B|^2)^2 - C |\psi_A|^2|\psi_B|^2 - \frac{\lambda^2}{2\epsilon}(\psi^\ast_A \psi^{\phantom\ast}_B + \psi^\ast_B \psi^{\phantom\ast}_A)^2. 
\end{align}
To find the minimum, we write:
\begin{align}
\psi_A = |\psi_A| e^{i\theta_A};
\quad
\psi_B = |\psi_B| e^{i\theta_B}.
\end{align}
Substituting the above in, we find:
\begin{align}
f = \frac{A}{2} (|\psi_A|^2+|\psi_B|^2) + \frac{B}{4}(|\psi_A|^2+|\psi_B|^2)^2 - C |\psi_A|^2|\psi_B|^2 - \frac{2\lambda^2}{\epsilon} |\psi_A|^2 |\psi_B|^2 \cos^2(\theta_A-\theta_B). 
\end{align}
When $A<0$, it is easy to see that the minima are given by:
\begin{align}
|\psi_A| = |\psi_B| = \sqrt{\frac{-A}{2(B-C-2\lambda^2/\epsilon)}},\quad \theta_A - \theta_B = 0,\pi,
\end{align}
which describe the states that the two N\'{e}el vectors are collinear.

For the minimal model used in the main text, we have shown that the phonon fluctuations in the early stages after the quench select the non-collinear magnetic order. We now show that the same is true in the Landau-Ginzburg model. The argument closely parallels that of the minimal model. Consider a domain of size $\xi$. Within the domain, the system has relaxed to its Landau energy minimum. In the limit of $\lambda\to 0$, the minima are given by:
\begin{align}
|\psi_A| = |\psi_B| = \sqrt{\frac{-A}{2(B-C)}} = \psi_0,
\end{align}
but the phase angles $\theta_A$ and $\theta_B$ are undetermined. We now consider the impact of phonons. We write:
\begin{align}
\psi_A(x) = \psi_0 e^{i (\theta_A + \delta \theta_A(x))};
\quad
\psi_B(x) = \psi_0 e^{i (\theta_B + \delta \theta_B(x))}.
\end{align}
Here, $\theta_{A,B}$ are the overall orientation of the N\'{e}el vector; $\delta\theta_{A,B} \sim O(Q)$ are the fluctuations due to the phonon fluctuations. Substituting the above into the Landau free energy density, and expanding to the quadratic order in $Q$, we find the total free energy is given by:
\begin{subequations}
\begin{align}
F = F_0+F_1 + F_2.
\end{align}
The zeroth order term is a constant:
\begin{align}
F_0 = -\frac{A^2}{4(B-C)}V;
\end{align}
$V$ is the volume of the domain. The first order term vanishes because $Q(x)$ is centered about $0$ at the early times:
\begin{align}
F_1 = - 2\lambda \psi^2_0 \cos(\theta_A-\theta_B) \int Q(x) d^3x = 0.
\end{align}
The second order term is given by:
\begin{align}
F_2 = \int \{\frac{K\psi^2_0}{2}  [(\nabla\delta\theta_A)^2+(\nabla\delta\theta_B)^2] + 2\lambda \psi^2_0 \sin(\theta_A-\theta_B)(\delta\theta_A(x)-\delta\theta_B(x)) Q(x)]\} d^3x.
\end{align}
\end{subequations}

In the momentum space,
\begin{align}
F = \sum_{q}\frac{K\psi^2_0}{2}  q^2(\delta\theta_{-qA}\delta\theta_{qA}+ \delta\theta_{-qB}\delta\theta_{qB}) + 2\lambda\psi^2_0 \sin(\theta_A-\theta_B)Q_{-q}(\delta\theta_{qA}-\delta\theta_{qB}).
\end{align}
Here we have omitted the physically unimportant constant term. Minimizing the above with respect to $\delta\theta_{qA}$ and $\delta\theta_{qB}$, we obtain:
\begin{align}
F = -\frac{1}{2}\sum_q \frac{8\lambda^2\psi^2_0}{Kq^2}\sin^2(\theta_A-\theta_B) Q_qQ_{-q}.
\end{align}
We may approximate $Q_q Q_{-q}$ by its expectation value:
\begin{align}
Q_q Q_{-q} \approx \overline{Q_qQ_{-q}} = \overline{Q^2_0},
\end{align}
where $\overline{Q^2_0}$ is the variance of the fluctuations. We thus find:
\begin{align}
F = -\frac{1}{2}  \frac{8\lambda^2\psi^2_0}{K}\sin^2(\theta_A-\theta_B)\times \sum_q \frac{1}{q^2} \propto -\sin^2(\theta_A-\theta_B) .
\end{align}
We see that the minima are at:
\begin{align}
\theta_A-\theta_B = \pm \frac{\pi}{2},
\end{align}
which correspond to the states that the two N\'{e}el vectors are orthogonal.

\subsection{Discretization and non-dimensionalization}

To facilitate numerical simulation, we apply the discretization and non-dimensionalization procedure to the Landau Ginzburg model similar to that of the minimal model. Here we simply quote the result. The Landau-Ginzburg free energy reads:
\begin{align}
F &= \frac{1}{2}\sum_{\langle ij\rangle} |\widetilde{\psi}_{iA}-\widetilde{\psi}_{jA}|^2 +  |\widetilde{\psi}_{iB}-\widetilde{\psi}_{jB}|^2 
\nonumber\\
& + \sum_i \frac{\widetilde{A}}{2} (|\widetilde{\psi}_{iA}|^2+|\widetilde{\psi}_{iB}|^2) + \frac{1}{4} (|\widetilde{\psi}_{iA}|^2+|\widetilde{\psi}_{iB}|^2)^2 - \widetilde{C}|\widetilde{\psi}_{iA}|^2 |\widetilde{\psi}_{iB}|^2
\nonumber\\
& + \sum_i -g \widetilde{Q}_i (\widetilde{\psi}^\ast_{iA} \widetilde{\psi}^{\phantom\ast}_{iB}+\widetilde{\psi}^\ast_{iB} \widetilde{\psi}^{\phantom\ast}_{iA}) + \frac{\widetilde{Q}^2_i}{2}.
\end{align}
The equations of motion reads:
\begin{align}
\dot{\widetilde{\psi}}_{iA} = -\frac{\partial \widetilde{F}}{\partial \widetilde{\psi}^\ast_{iA}} + \nu_{iA};
\quad
\dot{\widetilde{\psi}}_{iB} = -\frac{\partial \widetilde{F}}{\partial \widetilde{\psi}^\ast_{iB}} + \nu_{iB};
\quad
\dot{\widetilde{Q}}_{i} = -\frac{1}{\tau}\frac{\partial \widetilde{F}}{\partial \widetilde{Q}_{i}} + \eta_{i};
\end{align}
with white noises:
\begin{align}
\langle \nu_{iA}(\widetilde{t}) \nu_{jA}(\widetilde{t}') \rangle = \langle \nu_{iB}(\widetilde{t}) \nu_{jB}(\widetilde{t}') \rangle  = 2\widetilde{T}\delta_{ij}\delta(\widetilde{t}-\widetilde{t}').
\quad
\langle \eta_{i}(\widetilde{t}) \eta_{j}(\widetilde{t}') \rangle = \frac{2\widetilde{T}}{\tau}\delta_{ij}\delta(\widetilde{t}-\widetilde{t}').
\end{align}
Here, the parameter $\widetilde{T}$ is the dimensionless bath temperature. $\tau$ is the ratio of the phonon relaxation time to spin relaxation time. $\widetilde{g}$ is the dimensionless spin-phonon coupling. $\widetilde{C}>0$ controls the attraction between the two N\'{e}el orders. A sign change in $\widetilde{A}$ induces the phase transition to the ordered state. 

\subsection{Numerical results}

\begin{figure}
\includegraphics[width = 0.6\columnwidth]{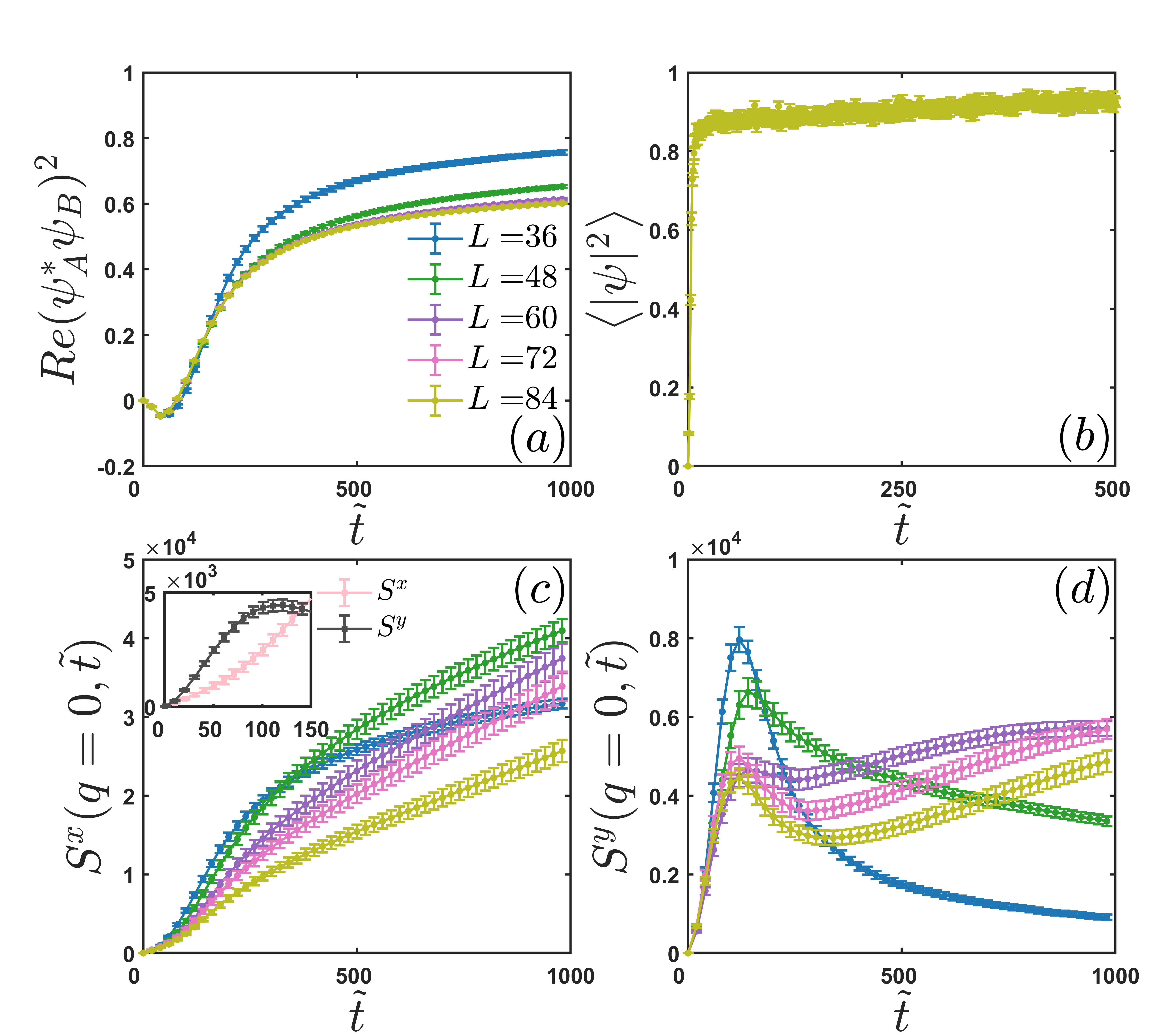}
\caption{Numerical simulation of the Landau-Ginzburg model. (a) The local spin anisotropy $\langle \mathrm{Re}(\psi^\ast_A \psi_B)^2\rangle$ as a function of dimensionless time $\widetilde{t}$. (b) The local amplitude of the order parameter $\langle |\psi|^2 \rangle$ as a function of time. (c) (d) The structure factor $S^x(q=0)$ and $S^y(q=0)$. The inset shows the comparison of the early time behavior.}
\label{fig:landau_model}
\end{figure}

In this subsection, we present the numerical results from a simulation of the Landau-Ginzburg model. We set the dimensionless coupling $g = 0.2$, the ratio of relaxation times $\tau = 150$, the bath temperature $\widetilde{T} = 0.1$. We choose $\widetilde{A} = -1.44$ and $\widetilde{C} = 0.2$.

For the initial conditions, we have a much larger degrees of freedom compared to that of the model used in the main text. Since our main purpose is to investigate the impact of the amplitude of the order parameter on the dynamics, we set the initial condition $\psi_{iA} = \psi_{iB} = 0$, i.e. the magnetic order is completely quenched locally. As for phonons, we use the same Gaussian ensemble with $\langle\widetilde{Q}^2\rangle = \widetilde{T}_i = 3$. This corresponds to setting $A\to \infty$ in the initial state.

Fig.~\ref{fig:landau_model} shows our results. We monitor the growth of the local amplitude of the order parameter by measuring the statistical average of $|\psi|^2$ \emph{for a fixed site}. We see that it grows rapidly in early time, and quickly saturates to $\sim 1$.This means that, as far as the physics of later time is concerned, we can fix the amplitude of the order parameter to a constant. 

We characterize the local spin anisotropy by $\mathrm{Re}(\psi^\ast_A \psi_B)^2$. If we set $\psi_A = \exp(i\theta_A)$ and $\psi_B = \exp(i\theta_B)$, then this quantity reduces to $\cos(2\theta_A-2\theta_B)$, coinciding with the quantity used in the main text. We find similar spin reorientation, i.e. the transition from a transient non-collinear configuration to collinear configuration (Fig.~\ref{fig:landau_model}a). 

We define the structure factors
\begin{subequations}
\begin{align}
S^x(\mathbf{q},\widetilde{t}) = \frac{1}{L^3}\sum_{i,j} \langle \mathrm{Re}(\psi^\ast_{iA}(\widetilde{t}) \psi^{\phantom\ast}_{iB}(\widetilde{t})) \mathrm{Re}(\psi^\ast_{jA}(\widetilde{t}) \psi^{\phantom\ast}_{jB}(\widetilde{t}))\rangle e^{-i\mathbf{q}\cdot(\mathbf{r}_i-\mathbf{r}_j)};
\\
S^y(\mathbf{q},\widetilde{t}) = \frac{1}{L^3}\sum_{i,j} \langle \mathrm{Im}(\psi^\ast_{iA}(\widetilde{t}) \psi^{\phantom\ast}_{iB}(\widetilde{t})) \mathrm{Im}(\psi^\ast_{jA}(\widetilde{t}) \psi^{\phantom\ast}_{jB}(\widetilde{t}))\rangle e^{-i\mathbf{q}\cdot(\mathbf{r}_i-\mathbf{r}_j)}.
\end{align}
\end{subequations}
This quantity reduces to that of $S^x$ used in the main text if we fix the amplitude of $\psi_A$ and $\psi_B$ to 1. We find their behaviors are qualitatively similar to that of the minimal model in the main text, and the two order parameter model in Sec.~\ref{sec:two_rotor}.

\section{Preliminary analysis of a triangular lattice spin-Peierls system\label{sec:zigzag}}

In this section, we analyze a model spin-Peierls system with triangular lattice geometry. Our model is inspired by the spin-Peierls instability in delafossite CuFeO$_2$~\cite{Kimura2006,Ye2006,Ye2007,Plumer2007,Wang2008,Quirion2009}. Our preliminary analysis of this model shows that the lattice distortion and the phonon fluctuations select different magnetic orders: the former favors the zigzag magnetic order as observed in equilibrium, whereas the latter favors the partial disordered state. Therefore, we expect that, when the phonon relaxation is slow, the initial phonon fluctuations inherited from the paramagnetic state would drive the system to the partial disoreded state, which gradually gives way to the zigzag magnetic order at late time as the phonons relax. 

We stress at the outset that our aim is to illustrate the diverse contexts in which the mechanisms revealed in our work could operate. A detailed investigation of the non-equilibrium dynamics of this materials is beyond the scope of present work.  

\subsection{Background}

\begin{figure}
\centering
\includegraphics[width = 0.8\textwidth]{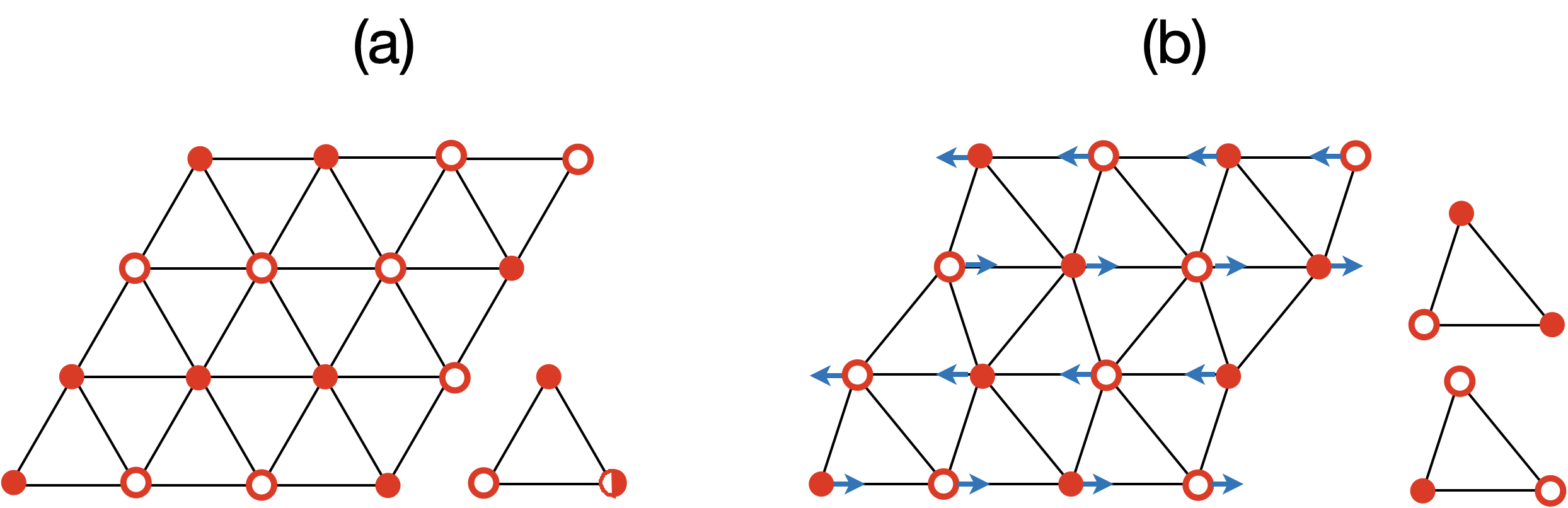}
\caption{(a) A degenerate ground state for Ising-like spins on ideal triangular lattice with nearest neighbor antiferromagnetic exchange interactions. Up and down spins are shown as filled and open circles. Inset: the degenerate ground states on an equilateral triangle. The half-filled circle indicates the spin orientation is undetermined. (b) A plausible explanation for the zigzag order observed in CuFeO$_2$ assumes spontaneous distortion of the lattice. The blue arrows indicate the postulated lattice distortion. Inset: ground states on an deformed triangle.}
\label{fig:zigzag_sketch}
\end{figure}

In zero magnetic field, the CuFeO$_2$ exhibits a four sublattice zigzag order as shown in Fig.~\ref{fig:zigzag_sketch}b. A heuristic microscopic understanding for the emergence of this order is the following~\cite{Wang2008}. Assuming the spins are Ising like, i.e. they are either parallel or anti-parallel to the common $c$ axis, the nearest neighbor antiferromagnetic exchange interaction results in geometric frustration. For each triangle, the exchange energy is minimal as long as two out of the three spins of the triangle are parallel and the third one is anti-parallel (Fig.~\ref{fig:zigzag_sketch}a, inset). Tiling the entire triangular lattice with these configurations result in a large amount of degenerate ground states whose number grows exponentially with increasing system size (Fig.~\ref{fig:zigzag_sketch}a). 

This massive degeneracy is lifted as soon as the lattice is allowed to relax. Consider the pattern of lattice site displacement shown in Fig.~\ref{fig:zigzag_sketch}b, where the horizontal chains slide in an staggered manner. This pattern corresponds to the condensation of a transverse acoustic phonon at an $M$ point of the Brillouin zone. By inspecting this pattern, we see that the triangles are no longer equilateral: one vertical nearest neighbor bond shrinks, and the other expands. Due to the spin-phonon coupling, the exchange interaction on the shrunk bond is stronger than the expanded bond. For a single triangle, this deformation selects a pair of configurations shown in the inset of Fig.~\ref{fig:zigzag_sketch}b. Tiling the entire plane with these two configurations results in the four sublattice zigzag order. 

The above argument assumes that the spins are Ising like. In CuFeO$_2$, the Ising anisotropy of the spins are likely to be moderate or weak. For isotropic exchange interactions, the antiferromagnetic exchange interaction is in fact not frustrated: the three spins on a triangle could form a $120^\circ$ order.  A careful analysis shows that the $120^\circ$ order gives way to the zigzag order when the strength of the magnitude of the spin phonon coupling exceeds a certain threshold~\cite{Wang2008}. Another missing ingredient of this explanation is that the zigzag order in CuFeO$_2$ features a periodic structure along the $c$ axis, which is not included in this purely two-dimensional picture. 

\subsection{Landau theory}

We build a Landau theory for the zigzag order in this system. From the preceding analysis, the essential ingredients are the order parameter for the zigzag order and the condensation of the transverse acoustic phonon at $M$ points. The zigzag order is described by:
\begin{align}
S(\mathbf{r}) = \sum_{i=1,2,3} (\psi_i e^{i\mathbf{Q}_i \cdot \mathbf{r}} + \psi^\ast_i e^{-i\mathbf{Q}_i\cdot \mathbf{r}}).
\end{align}
Here, $\mathbf{r}$ is the position vector of triangular lattice site. $\pm\mathbf{Q}_{1,2,3}$ are six characteristic wave vectors associated with the zigzag order. $\mathbf{Q}_1 = (\pi,0)$, and the other five vectors are obtained by $C_6$ rotation. $\psi_{1,2,3}$ are complex amplitudes. The zigzag order shown in Fig.~\ref{fig:zigzag_sketch}b corresponds to the amplitudes:
\begin{align}
\psi_{1} \propto \frac{1-i}{\sqrt{2}};\quad \psi_2 = \psi_3 = 0,
\end{align}
We take $\psi_{1,2,3}$ as the order parameters of the zigzag order. 

There are three inequivalent $M$ points in the Brillouin zone, on each residing an in-plane transverse acoustic phonon mode. We denote these modes by $u_{1,2,3}$.

We analyze the symmetry transformation properties of $\psi_{1,2,3}$ and $u_{1,2,3}$. $\psi_{1,2,3}$ and their complex conjugate form a star; together, they carry a six-dimensional irreducible representation of the two-dimensional wallpaper group $p6m$. Under time reversal, $\psi_{i} \to -\psi_i$. Under $C_2$ rotation, $\psi_i \to \psi^\ast_i$. Meanwhile, $u_{1,2,3}$ also form a star. Each $u_i$ carries the one-dimensional $B_2$ representation of the little co-group $D_2$ of the M point. Together, they carry a three-dimensional irreducible representation of the wallpaper group.

\begin{subequations}
The Landau energy density consists of three pieces:
\begin{align}
f = f_\mathrm{s} + f_\mathrm{ph}.
\end{align}
$f_\mathrm{s}$ is the Landau energy density for the zigzag order. $f_\mathrm{ph}$ is the coupling of the magnetic order to phonons.  

$f_\mathrm{s}$ is given by:
\begin{align}
f_\mathrm{s} = \frac{1}{2}\sum_i (K_\parallel |\nabla_\parallel \psi_i|^2+K_\perp |\nabla_\perp \psi_i|^2 + A |\psi_i|^2) + \frac{B}{4}(\sum_i |\psi_i|^2)^2 - C\sum_i |\psi_i|^4.
\label{eq:zigzag_fpsi}
\end{align}
As usual, $K_{\parallel,\perp}>0$, $B>0$. $A$ changes sign across the phase transition.  The first three terms are the only symmetry-allowed quadratic terms with at most two spatial derivatives. $\nabla_\parallel$ refers to the spatial gradient parallel to the wave vector $\mathbf{Q}_i$, whereas $\nabla_\perp$ refers to the gradient perpendicular to $\mathbf{Q}_i$. $B>0$ ensures the stability of the theory. $C>0$ reflects the repulsion between the order parameter at the three different $Q$ vectors. We note that there are additional symmetry-allowed quartic terms other than the one listed in $f_\mathrm{s}$; however, as these can be generated by phonons, we do not include them in $f_\mathrm{s}$.

We need to couple the phonons to the order parameters. We only consider coupling terms that is linear in phonon displacement $u_i$. Time reversal symmetry requires that the coupling must be at least quadratic in the order parameters. We note that $\psi^2_1$ and $(\psi^\ast_1)^2$ carries the same momentum as $u_1$, i.e. they are at $M_1$ point. Furthermore, the combination $i(\psi^2_1-(\psi^\ast_1)^2)$ carries the same $B_2$ representation of the little co-group $D_2$ as $u_1$ does (the imaginary unit is to ensure the term is real). The resulted term is thus given by:
\begin{align}
f_\mathrm{ph} = -i\lambda \sum_i u_i (\psi^2_i - (\psi^\ast_i)^2) +  \frac{\epsilon}{2}\sum_i u^2_i.
\end{align}
The coupling constant $\lambda>0$. $\epsilon>0$ is the elastic constant. We have modeled the zone boundary phonons as Einstein phonons for simplicity.
\end{subequations}

The above complete our description of the Landau theory.

\subsection{Order selection by lattice distortion}

We show that our theory correctly reproduces the magnetic orders discussed in the beginning of this section. It is sufficient for our purpose to set $\psi_i$ and $u_i$ to be spatially uniform. Minimizing the energy with respect to $u_i$ yields:
\begin{align}
f = \frac{A}{2}\sum_i |\psi_i|^2 + \frac{B}{4}(\sum_i |\psi_i|^2)^2 -C \sum_i |\psi_i|^4- \frac{\lambda^2}{2\epsilon}\sum_i [i(\psi^2_i - (\psi^\ast_i)^2)]^2.
\end{align}
To find the minimum of the above function, we write:
\begin{align}
\psi_i = |\psi_i| e^{i\theta_i}.
\end{align}
Substituting the above into $f$, we obtain:
\begin{align}
f = \frac{A}{2}\sum_i |\psi_i|^2 + \frac{B}{4}(\sum_i |\psi_i|)^2 -C\sum_i |\psi_i|^4 - \frac{2\lambda^2}{\epsilon}\sum_i |\psi|^4_i \sin^2 2\theta_i.
\end{align}
Minimizing $f$ with respect to $\theta_i$ yields:
\begin{align}
\theta_i = \pm \frac{\pi}{4}, \pm\frac{3\pi}{4}.
\end{align}
$f$ now reads:
\begin{align}
f = \frac{A}{2}\sum_i |\psi_i|^2 + \frac{B}{4}(\sum_i |\psi_i|^2)^2 -(C+ \frac{2\lambda^2}{\epsilon}) \sum_i |\psi_i|^4.
\end{align}
When $A<0$, the minima are:
\begin{align}
(|\psi_1|,|\psi_2|,|\psi_3|) = \sqrt{\frac{-A}{B-4C-8\lambda^2/\epsilon}}(1,0,0),
\end{align}
and the permutations of $|\psi_{1,2,3}|$.

\begin{figure}
\centering
\includegraphics[width = \columnwidth]{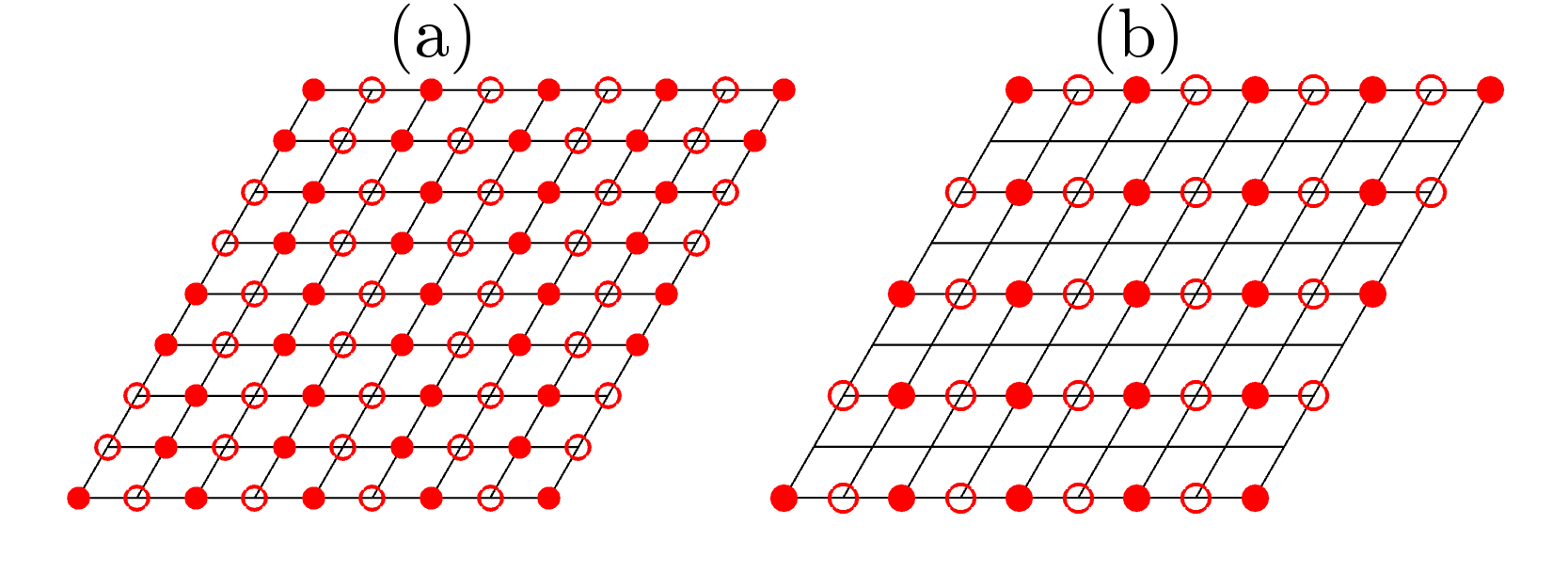}
\caption{(a) Zigzag order selected by lattice distortion. (b) Partially disordered state selected by the phonon fluctuations. The filled and open circles correspond to the up and down spins. The average magnetic moment is zero on empty sites, where the spins fluctuate between the up and down state.}
\label{fig:zigzag_phases}
\end{figure}

To summarize, we find in total 12 symmetry related ordered states:
\begin{align}
\psi_1 = \pm\sqrt{\frac{-A}{B-4C-8\lambda^2/\epsilon}} \frac{1\pm i}{\sqrt{2}}, \psi_2 = \psi_3 = 0;
\nonumber\\
\psi_2 = \pm\sqrt{\frac{-A}{B-4C-8\lambda^2/\epsilon}} \frac{1\pm i}{\sqrt{2}}, \psi_1 = \psi_3 = 0;
\nonumber\\
\psi_3 = \pm\sqrt{\frac{-A}{B-4C-8\lambda^2/\epsilon}} \frac{1\pm i}{\sqrt{2}}, \psi_1 = \psi_2 = 0.
\end{align}
They correspond to the 12 domains of the zigzag order, as shown in Fig.~\ref{fig:zigzag_phases}.

\subsection{Order selection by phonon fluctuations}

In the early stage after the quench, the phonons are symmetrically distributed about $u_i = 0$ due to the fluctuations inherited from the high temperature paramagnetic phase. In this subsection, we investigate the selection effect of the phonon fluctuations.

We consider a microscopically large but macroscopically small domain. Within the domain, the Landau energy has relaxed to its minimum. The problem is then finding this minimum for typical $u_i$ inside this domain. We do this by treating $\lambda$ as perturbation.

In the limit of $\lambda\to0$, the Landau energy minima are given by:
\begin{align}
|\psi_1| = \sqrt{\frac{-A}{B-4C}} = \psi_0 ; |\psi_2| = |\psi_3| = 0,
\end{align}
or the permutation of $|\psi_{1,2,3}|$. The phase angles are not determined. We now consider the case of small but finite $\lambda$. Without loss of generality, we write:
\begin{align}
\psi_1(x) = \psi_0 e^{i[\theta_1 + \delta\theta_1(x)]};
\quad
\psi_2(x) = \psi_3 = 0.
\end{align}
Here, $\delta\theta_1(x)$ is the fluctuations of the angle due to phonon fluctuations. Substituting the above into the Landau free energy, and expanding to quadratic order in $Q$, we find:
\begin{align}
F = F_0 + F_1 + F_2.
\end{align}
The zeroth order term is a constant:
\begin{align}
F_0 = -\frac{A^2}{4(B-4C)}V,
\end{align}
where $V$ is the volume of the domain. The first order term vanishes:
\begin{align}
F_1 = 2\lambda\psi^2_0 \sin(2\theta_1) \int u_1(x)d^2x = 0,
\end{align}
because $u_1(x)$ is symmetrically distributed about 0. The second order term reads:
\begin{align}
F_2 = \int \{\frac{K_\parallel \psi^2_0}{2} (\nabla_\parallel \delta\theta_1)^2 + \frac{K_\perp \psi^2_0}{2} (\nabla_\perp \delta\theta_1)^2 + 2\lambda\psi^2_0\cos2\theta_1 \delta\theta_1(x) u_1(x)\}d^2x.
\end{align}

In the momentum space, the free energy reads:
\begin{align}
F = \sum_q \frac{K_\parallel \psi^2_0 q^2_\parallel + K_\perp \psi^2_0 q^2_\perp }{2} \delta\theta_{-q1}\delta\theta_{q1} + 2\lambda\psi^2_0 \cos2\theta_1 u_{-q1} \delta\theta_{q1}.
\end{align}
Minimizing with respect to $\delta\theta_1$ yields:
\begin{align}
F = -\frac{1}{2}\sum_q \frac{4\lambda^2\psi^2_0}{K_\parallel q^2_\parallel + K_\perp q^2_\perp} u_{-q1}u_{q1}\cos^2 (2\theta_1) \approx -\frac{1}{2}\sum_q \frac{4\lambda^2\psi^2_0}{K_\parallel q^2_\parallel + K_\perp q^2_\perp} \overline{u^2_1} \cos^2\theta_1 \propto -\cos^2(2\theta_1).
\end{align}
In the second equality, we have replaced the product $u_{-q1}u_{q1}$ by its statistical average. The minima of the Landau free energy are:
\begin{align}
\theta_1 = 0, \frac{\pi}{2}, \pi, \frac{3\pi}{2}.
\end{align}
These states are visualized in Fig.~\ref{fig:zigzag_phases}b.

Comparing the case of lattice distortion with that of phonon fluctuations, we see that the difference lies in the phase of the order parameters. The former favors $\theta_i = n\pi/2 + \pi/4$, whereas the latter favors $\theta_i = n\pi/2$. The latter case correspond to the partial disordered state where half of the spins in the systems are fluctuating between up and down states.

\section{Preliminary analysis of a pyrochlore lattice spin-Peierls system}

In this section, we analyze a pyrochlore lattice model motivated by the spin-Peierls instability observed in chromate spinels ACr$_2$O$_4$, with A = Cd, Hg~\cite{Bergman2006,Matsuda2007,Matsuda2010}.  Our strategy is similar to Sec.~\ref{sec:zigzag}. We first construct a symmetry-constrained Landau theory for this system, and then analyze the selection of magnetic orders by lattice distortion and by phonon fluctuations. Similar to the case with triangular system, we find that the lattice distortion and the phonon fluctuations favor different magnetic orders. Therefore, by the same argument as in the main text and in Sec.~\ref{sec:zigzag}, the system would exhibit a transient magnetic order when the phonon relaxation is slow.

\subsection{Background}

\begin{figure}
\includegraphics[width=0.5\columnwidth]{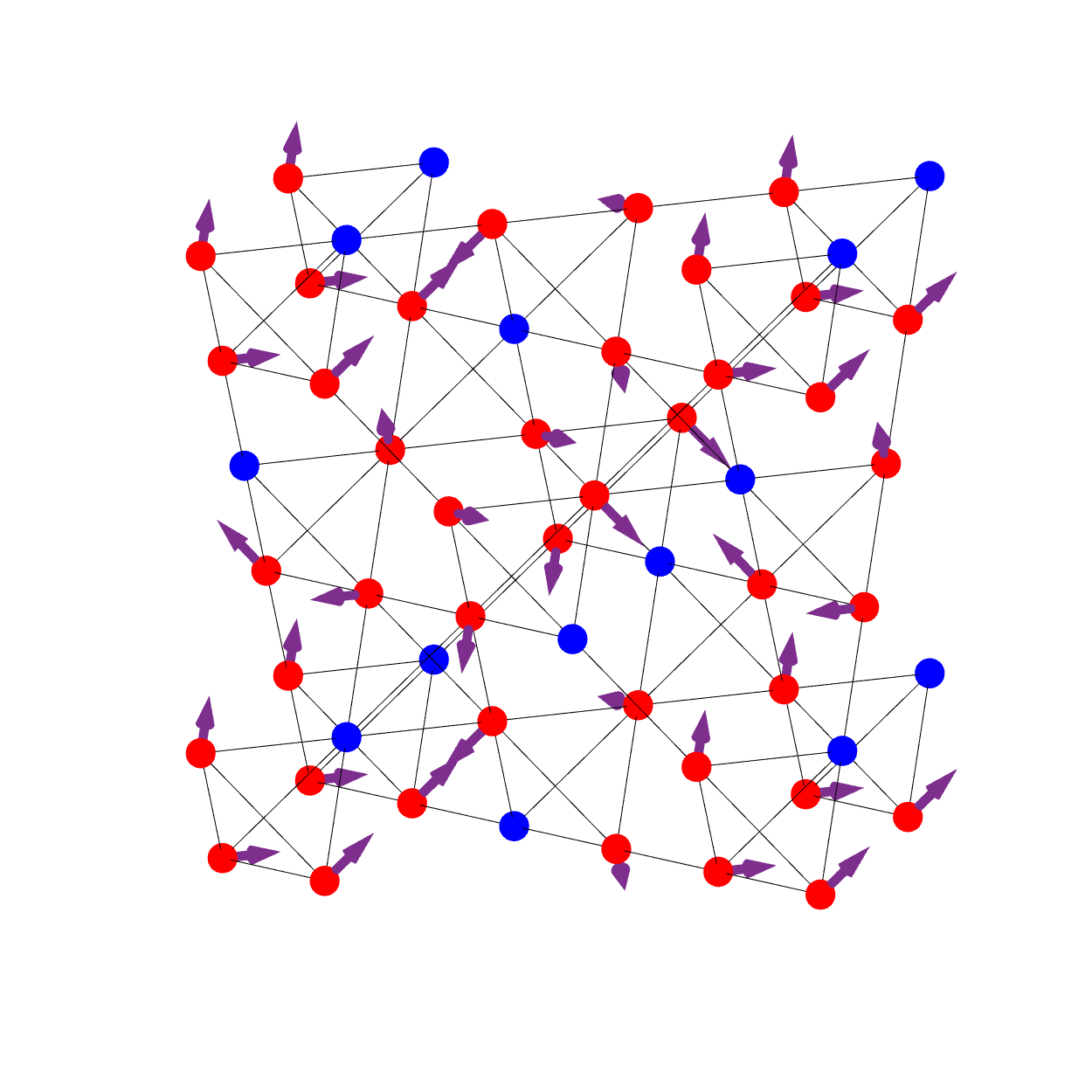}
\caption{The R state. Red and blue spheres respectively denote the majority and minority spins. Purple arrows show the displacement vectors of the lattice sites.}
\label{fig:pyro_R_state}
\end{figure}

The spinels ACr$_2$O$_4$ exhibits diverse magnetic orders at low temperature, accompanied by structure phase transitions. This is understood as a manifestation of the spin-Peierls instability as well as its subtle interplay with further neighbor exchange interactions in this family of materials.

While the zero magnetic field ground states are quite complex across this family, the situation is simplified somewhat in strong magnetic field. A sufficiently strong magnetic field drives the system to a half-magnetized plateau. Neutron scattering experiments suggest that the magnetic structure in the plateau state is described by the space group $P4_332$ for CdCr$_2$O$_4$ and HgCr$_2$O$_4$~\cite{Matsuda2007,Matsuda2010}. This finding is consistent with an earlier microscopic theory~\cite{Bergman2006}, which shows that the so-called R state is selected by complex spontaneous distortions of the pyrochlore lattice (Fig.~\ref{fig:pyro_R_state}).

We may heuristically understand the emergence of R state as follows. For the four spins belonging to a tetrahedron, three are parallel to the field (the majority spin), and one opposite to the field (the minority spin), producing in total four degenerate ground states. Tiling the pyrochlore lattice with the three-up-one-down states results in a massive degeneracy in the ground states, whose degree of degeneracy grows exponentially with the number of spins. 

This massive degeneracy is lifted by spontaneous lattice distortion. Inspecting the distortion pattern shown in Fig.~\ref{fig:pyro_R_state} reveals that, within each tetrahedron, there is always a vertex moving toward the opposing face. The exchange interactions on the shorter bonds are strengthened. The local ground state of the deformed tetrahedron is now unique: the minority spin must occupies the displaced vertex. Applying this rule to the whole lattice results in the R state. 

\subsection{Landau theory}

We construct the simplest Landau free energy that is consistent with the picture described above. We first need to identify the relevant order parameters. The magnetic Bragg peak of the R state is at the X points of the first Brillouin zone. The R state therefore must correspond to a certain irreducible representation of the space group $Fd\bar{3}m$ associated with the star $\{X_x, X_y, X_z\}$. $X_x = (100)$, $X_y = (010)$, and $X_z = (001)$. For each $X_i$, there are four inequivalent, two-dimensional small representations, labeled as $E_{1,2,3,4}$. The R state is described by the small representation $E_2$. We write the spin density modulation as:
\begin{align}
\delta S (r) = \sum_{i=x,y,z}\sum_{\alpha=1,2} \psi_{i\alpha} \Psi_{i\alpha}(r).
\end{align}
$\psi_{i\alpha}$ are real order parameters. $\Psi_{i\alpha}(r)$ are real basis functions that carry the said small representation at momentum $X_i$. They are given by:
\begin{subequations}
\begin{align}
\Psi_{x1}(r) = e^{-i\frac{\pi}{4}} (\Psi_{\mathbf{X}_y A} - \Psi_{\mathbf{X}_y B}) + e^{i\frac{\pi}{4}} (\Psi_{\mathbf{X}_y D} - \Psi_{\mathbf{X}_y C}); \\
\Psi_{x2}(r) = e^{-i\frac{\pi}{4}} (\Psi_{\mathbf{X}_y A} - \Psi_{\mathbf{X}_y B}) - e^{i\frac{\pi}{4}} (\Psi_{\mathbf{X}_y D} - \Psi_{\mathbf{X}_y C}); \\
\Psi_{y1}(r) = e^{-i\frac{\pi}{4}} (\Psi_{\mathbf{X}_y A} - \Psi_{\mathbf{X}_y C}) + e^{i\frac{\pi}{4}} (\Psi_{\mathbf{X}_y B} - \Psi_{\mathbf{X}_y D});  \\
\Psi_{y2}(r) = e^{-i\frac{\pi}{4}} (\Psi_{\mathbf{X}_y A} - \Psi_{\mathbf{X}_y C}) - e^{i\frac{\pi}{4}} (\Psi_{\mathbf{X}_y B} - \Psi_{\mathbf{X}_y D}); \\
\Psi_{z1}(r) = e^{-i\frac{\pi}{4}} (\Psi_{\mathbf{X}_y A} - \Psi_{\mathbf{X}_y D}) + e^{i\frac{\pi}{4}} (\Psi_{\mathbf{X}_y C} - \Psi_{\mathbf{X}_y B}); \\
\Psi_{z2}(r) = e^{-i\frac{\pi}{4}} (\Psi_{\mathbf{X}_y A} - \Psi_{\mathbf{X}_y D}) - e^{i\frac{\pi}{4}} (\Psi_{\mathbf{X}_y C} - \Psi_{\mathbf{X}_y B}).
\end{align}
\end{subequations}
Here, we have defined the Bloch state:
\begin{align}
\Psi_{\mathbf{Q}\alpha} (r) = \left\{ \begin{array}{cc}
e^{i\mathbf{Q}\cdot r} & (r\in \alpha) \\
0 & (\mathrm{otherwise})
\end{array}\right. ,
\end{align}
i.e. it is a plane wave in sublattice $\alpha$. We use the convention that the four sublattice sites are located at $r_A = [\frac{1}{8}\frac{1}{8}\frac{1}{8}]$, $r_B = [\frac{1}{8}\frac{\bar{1}}{8}\frac{\bar{1}}{8}]$, $r_C = [\frac{\bar{1}}{8}\frac{1}{8}\frac{\bar{1}}{8}]$, and $r_D = [\frac{\bar{1}}{8}\frac{\bar{1}}{8}\frac{1}{8}]$ relative to the center of a tetrahedron.

The lattice distortion corresponds to the condensation of transverse acoustic phonons at $X$ points that also carry the $E_2$ small representation. They are described by the phonon fields, $u_{i\alpha}$, where $i$ labels the three X points, and $\alpha$ the two basis functions associated with a given $X_i$.

We are now ready to write down the Landau free energy:
\begin{subequations}
\begin{align}
f = f_\mathrm{s} + f_\mathrm{ph}.
\end{align}
The magnetic part reads:
\begin{align}
f_\mathrm{s} = \frac{1}{2}\sum_{i\alpha} [K_\parallel (\nabla_\parallel \psi_{ i\alpha})^2+K_\perp (\nabla_\perp \psi_{i\alpha})^2 + A (\psi_{i\alpha})^2] + \frac{B}{4}(\sum_{i\alpha} \psi^2_{i\alpha})^2 - C\sum_\alpha \psi_{x\alpha}\psi_{y\alpha}\psi_{z\alpha} + C'\sum_{i\alpha} \psi^4_{i\alpha}.
\end{align}
Here, $K_\parallel,K_\perp>0$ control the stiffness parallel with (perpendicular to) $X_\alpha$. The sign change in $A$ induces condensation of $\psi_{i\alpha}$. $B>0$ ensures the stability of the theory. Note there are additional quartic terms that are allowed by symmetry. However, they can be generated by phonons, and, therefore, are dropped from $f_\mathrm{s}$. We may set $C>0$ without loss of generality. The sign of $C'$ is not determined a priori.

The phonon part reads:
\begin{align}
f_\mathrm{ph} = -\lambda\sum_\alpha (\psi_{x\alpha}\psi_{y\alpha}u_{z\alpha} + \psi_{y\alpha}\psi_{z\alpha}u_{x\alpha} + \psi_{z\alpha}\psi_{x\alpha}u_{y\alpha}) + \frac{\epsilon}{2}\sum_{i\alpha} u^2_{i\alpha},
\end{align}
\end{subequations}
where $\lambda>0$ is the coupling constant. $\epsilon>0$ is the stiffness.

\subsection{Selection by lattice distortion}

We first consider the selection of magnetic orders by lattice distortion. To this end, we assume $\psi_{i\alpha}$ are all spatially uniform. Minimizing with respect to $u_{i\alpha}$ yields:
\begin{align}
f = \frac{A}{2}\sum_{i\alpha}  (\psi_{i\alpha})^2 + \frac{B}{4}(\sum_{i\alpha} \psi^2_{i\alpha})^2 - C\sum_\alpha \psi_{x\alpha}\psi_{y\alpha}\psi_{z\alpha} + C'\sum_{i\alpha} \psi^4_{i\alpha} - \frac{\lambda^2}{2\epsilon}\sum_{\alpha} (\psi^2_{x\alpha} \psi^2_{y\alpha}+\psi^2_{y\alpha} \psi^2_{z\alpha}+\psi^2_{z\alpha} \psi^2_{x\alpha}).
\end{align}
We see that the phonons also generated a quartic term that can compete against/collaborate with $C$ and $C'$. To simplify the analytical calculation, we assume the $B$ term, which exhibits $O(6)$ rotational symmetry, is the dominant interaction term. This $O(6)$ symmetry is then broken by the remaining terms $C$, $C'$,  $\lambda^2/(2\epsilon)$. 

Minimizing the first two terms yield the condition:
\begin{align}
\sum_{i\alpha} \psi^2_{i\alpha} = \frac{-A}{B} = \psi^2_0.
\end{align}
This condition describes a 5-sphere; we write $\psi_{i\alpha} = \psi_0 n_{i\alpha}$, where $n$ is a 6-dimensional unit vector. Substituting the above into the expression for Landau energy, we obtain:
\begin{align}
f = -\psi^3_0 C\sum_\alpha n_{x\alpha} n_{y\alpha} n_{z\alpha} + \psi^4_0 C'\sum_{i\alpha} n^4_{i\alpha} - \Delta \sum_{\alpha} n^2_{x\alpha} n^2_{y\alpha}+n^2_{y\alpha} n^2_{z\alpha}+n^2_{z\alpha} n^2_{x\alpha} + \mathrm{Const}.
\end{align}
Here, $\Delta = \psi^4_0 \lambda^2/(2\epsilon)$. 

The remaining task is to find the minimum of $f$. We do this in two steps. We begin by writing:
\begin{subequations}
\begin{align}
f = \sum_\alpha f_\alpha,
\end{align}
where
\begin{align}
f_\alpha = -\psi^3_0 C n_{x\alpha} n_{y\alpha} n_{z\alpha} + \psi^4_0 C' \sum_{i} n^4_{i\alpha} - \Delta (n^2_{x\alpha} n^2_{y\alpha}+n^2_{y\alpha} n^2_{z\alpha}+n^2_{z\alpha} n^2_{x\alpha}).
\end{align}
\end{subequations}
In the first step, we minimize $f_\alpha$ individually with the constraint $\sum_i n^2_{i\alpha} = l^2_\alpha$. To do this, we massage $f_\alpha$:
\begin{align}
f_\alpha = -\psi^3_0 C n_{x\alpha} n_{y\alpha} n_{z\alpha} + \psi^4_0 C' \sum_{i} n^4_{i\alpha} - \frac{\Delta}{2} (l^4_\alpha -\sum_i n^4_{i\alpha}) 
\nonumber\\
= -\psi^3_0 C n_{x\alpha} n_{y\alpha} n_{z\alpha} + (\psi^4_0 C' + \frac{\Delta}{2} ) \sum_{i} n^4_{i\alpha} + \mathrm{Const}.
\end{align}
If $\psi^4_0C'+\Delta/2>0$, then the minima of $f_\alpha$ are located at:
\begin{align}
(n_{x\alpha},n_{y\alpha},n_{z\alpha}) = \frac{l_\alpha}{\sqrt{3}}\left\{ \begin{array}{c}
(1,1,1) \\
(1,-1,-1) \\
(-1,1,-1) \\
(-1,-1,1)
\end{array}\right. ;
\quad
f_\alpha = -\psi^3_0 C \frac{l^3_\alpha}{3\sqrt{3}} + (\psi^4_0 C' - \Delta) \frac{l^4_\alpha}{3}.
\end{align}
If $\psi^4_0C'+\Delta/2<0$, then the minima of $f_\alpha$ are located at:
\begin{align}
(n_{x\alpha},n_{y\alpha},n_{z\alpha}) = l_\alpha \left\{ \begin{array}{c}
(\pm1,0,0) \\
(0,\pm1,0) \\
(0,0,\pm1) 
\end{array}\right. ;
\quad
f_\alpha = \psi^4_0 C' \frac{l^4_\alpha}{3}.
\end{align}

In the second step, we minimize the sum $f = \sum_\alpha f_\alpha$ with respect to $l_\alpha$ subject to the constraint $\sum_\alpha l^2_\alpha = 1$ and $l_\alpha>0$. In the case of $\psi^4_0C'+\Delta/2>0$, we have:
\begin{align}
f = \sum_\alpha [-\psi^3_0 C  \frac{l^3_\alpha}{3\sqrt{3}} + (\psi^4_0 C' - \Delta) \frac{l^4_\alpha}{3}].
\end{align}
The minima are at:
\begin{align}
(l_1,l_2) =  \left\{ \begin{array}{cc}
(1,0),(0,1) & -(2-\sqrt{2})\psi^3_0C + \sqrt{3}(\psi^4_0C'-\Delta)<0 \\
(1/\sqrt{2},1/\sqrt{2}) & -(2-\sqrt{2})\psi^3_0C + \sqrt{3}(\psi^4_0C'-\Delta)>0
\end{array} \right. .
\end{align}
In the case of $\psi^4_0C'+\Delta/2<0$, we have:
\begin{align}
f = \sum_\alpha \psi^4_0 C' \frac{l^4_\alpha}{3}.
\end{align}
The minima are at:
\begin{align}
(l_1,l_2) =  \left\{ \begin{array}{cc}
(1,0),(0,1) & C'<0 \\
(1/\sqrt{2},1/\sqrt{2}) & C'>0
\end{array} \right. .
\end{align}

\begin{figure}
\includegraphics[width = \columnwidth]{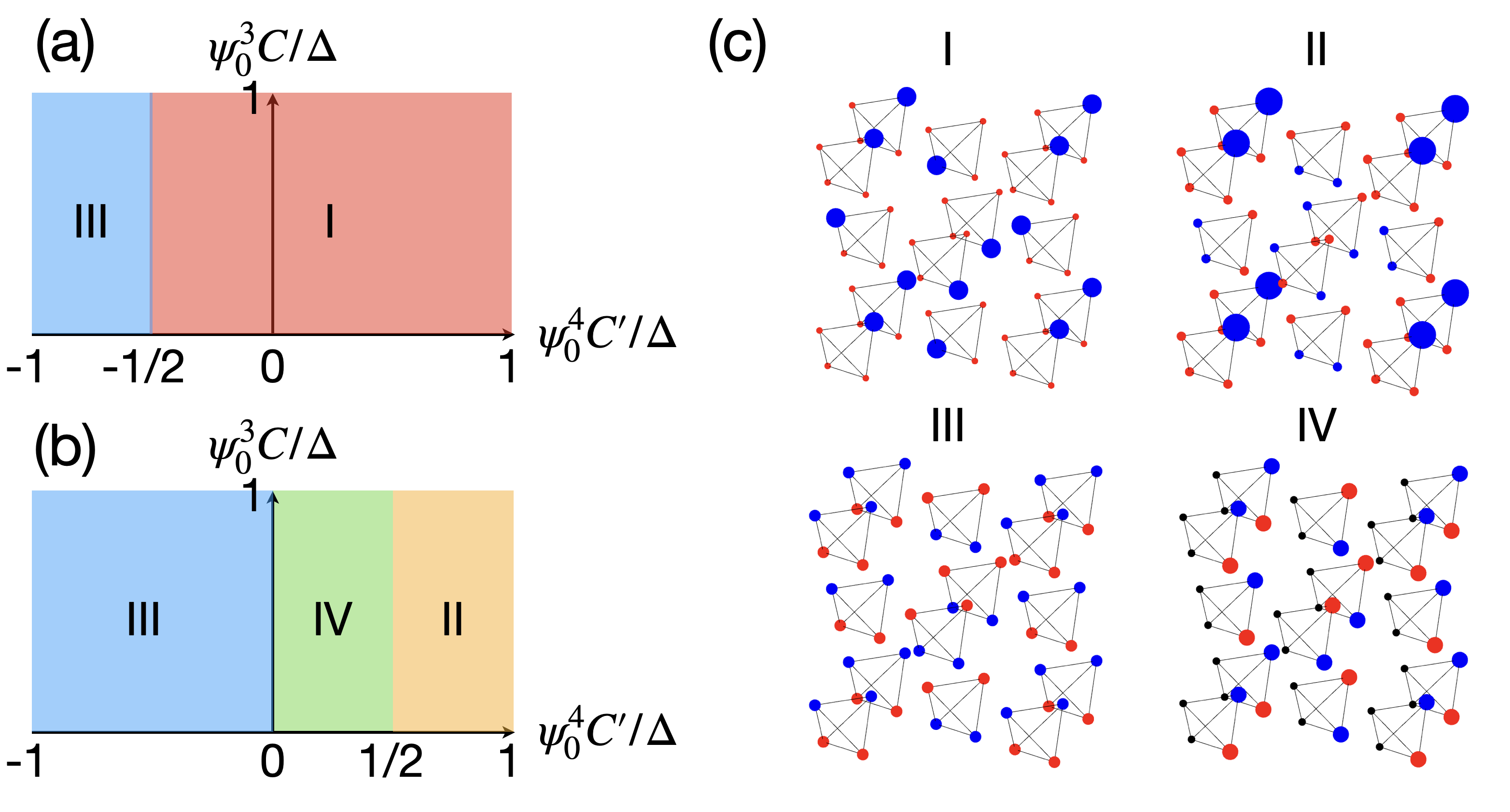}
\caption{(a) Magnetic order selection due to the interplay between the lattice distortion and the other interaction terms. Roman numerals label the different phases defined in the main text. (b) Similar to (a) but for the case of phonon fluctuations. (c) Visualization of the spin modulation in the four competing magnetic orders. Red and blue spheres correspond to the positive and negative modulation of the magnetization. The radius is proportional to the modulation magnitude. Sites with zero modulation is colored in black. For clarity, only half of the tetrahedra (the ``up" tetrahedra) are shown.}
\label{fig:pyro_phase_diag}
\end{figure}

In summary, we identify four potential phases in this model, dubbed phase I, II, III, and IV. In particular, the phase I is identical to the R state. A visualization of these phases are shown in Fig.~\ref{fig:pyro_phase_diag}c.
\begin{subequations}
\begin{align}
\begin{pmatrix}
\psi_{x1} \\
\psi_{y1} \\
\psi_{z1} \\
\psi_{x2} \\
\psi_{y2} \\
\psi_{z2}
\end{pmatrix} = \frac{\psi_0}{\sqrt{3}} \times \begin{pmatrix}
1 \\
1 \\
1 \\
0 \\
0 \\
0
\end{pmatrix},\, \begin{pmatrix}
1 \\
-1 \\
-1 \\
0 \\
0 \\
0
\end{pmatrix},\, \begin{pmatrix}
-1 \\
1 \\
-1 \\
0 \\
0 \\
0
\end{pmatrix},\, \begin{pmatrix}
-1 \\
-1 \\
1 \\
0 \\
0 \\
0
\end{pmatrix}, \, \begin{pmatrix}
0 \\
0 \\
0 \\
1 \\
1 \\
1
\end{pmatrix}, \, \begin{pmatrix}
0 \\
0 \\
0 \\
1 \\
-1 \\
-1
\end{pmatrix}, \, \begin{pmatrix}
0 \\
0 \\
0 \\
-1 \\
1 \\
-1
\end{pmatrix}, \, \begin{pmatrix}
0 \\
0 \\
0 \\
-1 \\
-1 \\
1
\end{pmatrix}. \, (\textrm{Phase I}).
\end{align}

\begin{align}
\begin{pmatrix}
\psi_{x1} \\
\psi_{y1} \\
\psi_{z1} \\
\psi_{x2} \\
\psi_{y2} \\
\psi_{z2}
\end{pmatrix} = \frac{\psi_0}{\sqrt{6}} \times \begin{pmatrix}
1 \\
1 \\
1 \\
1 \\
1 \\
1
\end{pmatrix},\, \begin{pmatrix}
1 \\
-1 \\
-1 \\
1 \\
-1 \\
-1
\end{pmatrix},\, \begin{pmatrix}
-1 \\
1 \\
-1 \\
-1 \\
1 \\
-1
\end{pmatrix},\, \begin{pmatrix}
-1 \\
-1 \\
1 \\
-1 \\
-1 \\
1
\end{pmatrix}. \, (\textrm{Phase II}).
\end{align}

\begin{align}
\begin{pmatrix}
\psi_{x1} \\
\psi_{y1} \\
\psi_{z1} \\
\psi_{x2} \\
\psi_{y2} \\
\psi_{z2}
\end{pmatrix} = \psi_0 \times \begin{pmatrix}
\pm1 \\
0 \\
0 \\
0 \\
0 \\
0
\end{pmatrix},\, \begin{pmatrix}
0 \\
\pm1 \\
0 \\
0 \\
0 \\
0
\end{pmatrix},\, \begin{pmatrix}
0 \\
0 \\
\pm1 \\
0 \\
0 \\
0
\end{pmatrix},\, \begin{pmatrix}
0 \\
0 \\
0 \\
\pm1 \\
0 \\
0
\end{pmatrix}, \, \begin{pmatrix}
0 \\
0 \\
0 \\
0 \\
\pm1 \\
0
\end{pmatrix}, \, \begin{pmatrix}
0 \\
0 \\
0 \\
0 \\
0 \\
\pm1
\end{pmatrix}. \, (\textrm{Phase III}).
\end{align}

\begin{align}
\begin{pmatrix}
\psi_{x1} \\
\psi_{y1} \\
\psi_{z1} \\
\psi_{x2} \\
\psi_{y2} \\
\psi_{z2}
\end{pmatrix} = \frac{\psi_0}{\sqrt{2}} \times \begin{pmatrix}
\pm1 \\
0 \\
0 \\
\pm1 \\
0 \\
0
\end{pmatrix},\, \begin{pmatrix}
0 \\
\pm1 \\
0 \\
0 \\
\pm1 \\
0
\end{pmatrix},\, \begin{pmatrix}
0 \\
0 \\
\pm1 \\
0 \\
0 \\
\pm1
\end{pmatrix}. \, (\textrm{Phase IV}).
\end{align}
\end{subequations}

The resulted phase diagram is shown in Fig.~\ref{fig:pyro_phase_diag}a. We focus on the window with $\psi^3_0 C<\Delta$ and $\psi^4_0|C'|<\Delta$, i.e. the regime where the $C$ and $C'$ terms can potentially compete with the lattice distortion. The physics outside this window is beyond the scope of this analysis in that the lattice distortion would be a sub-leading effect. We see that the phase I, i.e. the R state, occupies a large portion of the window, consistent with the experimental results.

\subsection{Selection by phonon fluctuations}

We then consider the selection of magnetic orders by phonon fluctuations. Same as before, we assume $B$ is the dominant interaction scale. We thus may treat the selection effect of various terms in the Landau theory separately. 

We first derive an effective Landau energy function by ``integrating out" phonon fluctuations. For now, we set $C$ and $C'$ to zero. We consider a domain, within which the Landau energy has reached its minimum. We write:
\begin{align}
\psi_{i\alpha}(x) = \psi_0[n_{i\alpha}(x) + \delta n_{i\alpha}(x)].
\end{align} 
where $\psi_0 = \sqrt{-A/B}$ is the amplitude of the order parameter. $n_{i\alpha}$ is the average orientation of the order parameter in the six-dimensional space, and $\delta n_{i\alpha}(x)$ is the fluctuation due to phonon.  The normalization of $n$ vector induces the following linear constraint:
\begin{align}
\sum_{i\alpha} n_{i\alpha} \delta n_{i\alpha}(x) = 0.
\end{align}
Similar to the treatment in Sec.~\ref{sec:zigzag}, we expand the total Landau energy to quadratic order in $\delta n_{i\alpha}$:
\begin{align}
F = \frac{\psi^2_0}{2}\sum_{i\alpha} [K_\parallel (\nabla_\parallel \delta n_{ i\alpha})^2+K_\perp (\nabla_\perp \delta n_{i\alpha})^2] - \psi^2_0 \lambda \sum_{\alpha} (n_{x\alpha} \delta n_{y\alpha} u_{z\alpha}+\textrm{permutations of $x,y,z$}) + \mathrm{Const}.
\end{align}
In the momentum space,
\begin{align}
F = \frac{\psi^2_0}{2} \sum_{q,i\alpha} (K_\parallel q^2_\parallel+K_\perp q^2_\perp)\delta n_{i\alpha}(-q)\delta n_{i\alpha}(q)- \psi^2_0 \lambda \sum_{\alpha} (n_{x\alpha} u_{z\alpha} (-q) \delta n_{y\alpha}(q) +\textrm{permutations of $x,y,z$})
\nonumber\\
= \frac{\psi^2_0}{2} \sum_{q,i\alpha} (K_\parallel q^2_\parallel+K_\perp q^2_\perp)\delta n_{i\alpha}(-q)\delta n_{i\alpha}(q)- \psi^2_0 \lambda \sum_{i\alpha}  j_{i\alpha}(-q) \delta n_{i\alpha}(q).
\end{align}
Here, we have defined:
\begin{align}
j_{x\alpha}(q) = n_{y\alpha} \delta u_{z\alpha}(q) +  n_{z\alpha} \delta u_{y\alpha}(q);
\nonumber\\
j_{y\alpha}(q) = n_{z\alpha} \delta u_{x\alpha}(q) +  n_{x\alpha} \delta u_{z\alpha}(q);
\nonumber\\
j_{z\alpha}(q) = n_{x\alpha} \delta u_{y\alpha}(q) +  n_{y\alpha} \delta u_{x\alpha}(q).
\end{align}
The constraint reads:
\begin{align}
\sum_{i\alpha} n_{i\alpha} \delta n_{i\alpha}(q) = 0.
\end{align}

Minimizing $F$ with respect to $\delta n_{i\alpha}$ yields:
\begin{align}
F = -\frac{\psi^2_0 \lambda^2}{2}\sum_{q} \frac{1}{K_\parallel q^2_\parallel + K_\perp q^2_\perp} [\sum_{i\alpha} j_{i\alpha}(-q)j_{i\alpha}(q) - \sum_{i\alpha} n_{i\alpha} j_{i\alpha}(-q) \sum_{i'\alpha'}n_{i'\alpha'} j_{i'\alpha'}(q)].
\end{align} 
We replace the product of $j$ by their statistical averages:
\begin{subequations}
\begin{align}
\sum_{i\alpha} \overline{j_{i\alpha}(-q)j_{i\alpha}(q)} = 2\overline{u^2_0} \sum_\alpha (n^2_{x\alpha}+n^2_{y\alpha}+n^2_{z\alpha}) = 4\overline{u^2_0};\\
\overline{\sum_{i\alpha} n_{i\alpha} j_{i\alpha}(-q) \sum_{i'\alpha'}n_{i'\beta'} j_{i'\beta'}(q)} = 4\overline{u^2_0} \sum_\alpha n^2_{x\alpha}n^2_{y\alpha} + n^2_{y\alpha}n^2_{z\alpha} + n^2_{z\alpha}n^2_{x\alpha}.
\end{align}
\end{subequations}
Substituting the above in, we find:
\begin{align}
F = 2\psi^2_0 \lambda^2 \overline{u^2_0} \sum_{q} \frac{1}{K_\parallel q^2_\parallel + K_\perp q^2_\perp} \sum_\alpha n^2_{x\alpha}n^2_{y\alpha} + n^2_{y\alpha}n^2_{z\alpha} + n^2_{z\alpha}n^2_{x\alpha}.
\end{align}
Therefore, the phonon fluctuations generate an effective interaction term in the Landau free energy density:
\begin{align}
f = \Delta \sum_\alpha n^2_{x\alpha}n^2_{y\alpha} + n^2_{y\alpha}n^2_{z\alpha} + n^2_{z\alpha}n^2_{x\alpha},
\end{align}
where
\begin{align}
\Delta =  \frac{2 \psi^2_0\lambda^2 \overline{u^2_0}}{V}\sum_{q} \frac{1}{K_\parallel q^2_\parallel + K_\perp q^2_\perp} >0.
\end{align}

We are in position to investigate the interplay between the selection effects of the phonon fluctuations vis a vis the $C$ and $C'$ terms. The Landau free energy reads:
\begin{align}
f = -\psi^3_0 C\sum_\alpha n_{x\alpha} n_{y\alpha} n_{z\alpha} + \psi^4_0 C'\sum_{i\alpha} n^4_{i\alpha} + \Delta \sum_{\alpha} (n^2_{x\alpha} n^2_{y\alpha}+n^2_{y\alpha} n^2_{z\alpha}+n^2_{z\alpha} n^2_{x\alpha}).
\end{align} 
We use the same method as the previous subsection to determine the minima of $f$. The resulted phase diagram is shown in Fig.~\ref{fig:pyro_phase_diag}b.

We now compare the magnetic order selection in the two cases. Assuming the system is in the R state (phase I) in equilibrium, i.e. $\psi^4_0C'/\Delta>1/2$, the magnetic orders selected by the phonon fluctuations are the phase II, III, and IV. We thus conclude that the phonon fluctuations will favor transient magnetic orders that are different from the equilibrium one.

\section{Transient selection of competing order with underdamped phonons}

We assume overdamped dynamics for simplicity; it by no means implies that the result is only valid in that limit. We choose the overdamped dynamics in the same spirit as the Landau-Ginzburg theory, where the dynamics is often assumed to be relaxational. 

In this section, we show that the transient selection of competing magnetic order is robust if the phonons are underdamped. To this end, we consider the following model:
\begin{align}
H=-J \sum_{\langle ij \rangle}\cos(\theta_i-\theta_j)-g\sum_i Q_i \cos\theta_i+\sum_i\frac{P_i^2}{2m}+\frac{m\omega^2 Q_i^2 }{2}.
\end{align}
Compared with the model in the main text, we have explicitly introduced the momentum $P_i$ conjugate to the phonon displacement $Q_i$. The equations of motion read:
\begin{subequations}
\begin{align}
\dot{\theta_i} &=-\gamma_{\theta}\frac{\partial H}{\partial \theta_i}+\eta_{\theta} =-\gamma_{\theta}( J\sum_{j\in N_i}\sin{(\theta_i-\theta_j)}+g\sin\theta_i Q_i )+\eta_{\theta};
\\
\dot{Q}_i &= \frac{\partial H}{\partial P_i}=\frac{P_i}{m};
\\
\dot{P}_i &=-\frac{\partial H}{\partial Q_i}-\gamma_P P_i+\eta_p=-m\omega^2Q_i+g\cos{\theta_i}-\gamma_P P_i+\eta_p.
\end{align}
Here, $\eta_\theta$ and $\eta_P$ are Gaussian white noise:
\begin{align}
\langle\eta_{\theta}(t)\eta_{\theta}(t^{\prime})\rangle=2\gamma_{\theta} k_BT\delta(t-t^{\prime}),
\quad
\langle\eta_{p}(t)\eta_{p}(t^{\prime})\rangle=2\gamma_Pm k_BT\delta(t-t^{\prime}).
\end{align}
\end{subequations}
$k_B T$ is the bath temperature. $\gamma_P$ is the damping of the oscillator; when $\gamma_P<\omega$, the phonon dynamics is underdamped.

It is convenient to render all variables dimensionless by appropriate rescaling:
\begin{align}
\bar{t}=\gamma_{\theta}J t,\quad 
\bar{P}=\frac{P}{\sqrt{mJ}},\quad  
\bar{Q}=Q\sqrt{mJ}\gamma_{\theta},\quad
\bar{\eta}_{\theta}=\frac{\eta_{\theta}}{\gamma_{\theta}J},\quad 
\bar{\eta}_{p}=\frac{\eta_{p}}{\sqrt{mJ}\gamma_{\theta}J}.
\end{align}
The dimensionless coupling constant, phonon frequency, phonon damping rate, and temperature are defined as:
\begin{align}
\bar{g}=\frac{g}{\sqrt{mJ}\gamma_{\theta}J},\quad
\bar{\omega}=\frac{\omega}{\gamma_{\theta}J},\quad
\bar{\gamma}_{P}=\frac{\gamma_{P}}{\gamma_{\theta}J},\quad
\bar{T}=\frac{k_BT}{J}.
\end{align}
The equations of motion now assume the following form:
\begin{subequations}
\begin{align}
\frac{\partial {\theta}_i }{\partial \bar{t}} &=
-\sum_{j\in N_i}\sin{(\theta_i-\theta_j)}-\bar{g}\sin\theta_i \bar{Q}_i+\bar{\eta}_{\theta};
\\
\frac{\partial \bar{Q}_i }{\partial \bar{t}} &= \bar{P}_i;
\\
\frac{\partial \bar{P}_i }{\partial \bar{t}} &= -\bar{\omega}^2 \bar{Q}_i+\bar{g}\cos{\theta_i}-\bar{\gamma}_P \bar{P}_i+\bar{\eta}_p;
\end{align}
with 
\begin{align}
\langle \overline{\eta}_{\theta}(\overline{t}) \overline{\eta}_{\theta}(\overline{t}')\rangle=2\overline{T}\delta(\overline{t}-\overline{t}'),
\quad
\langle \overline{\eta}_{p}(\overline{t}) \overline{\eta}_{p}(\overline{t}')\rangle=2\overline{\gamma}_P \overline{T} \delta(\overline{t}-\overline{t}').
\end{align}
\end{subequations}

\begin{figure}
\centering
\includegraphics[width = 0.6\textwidth]{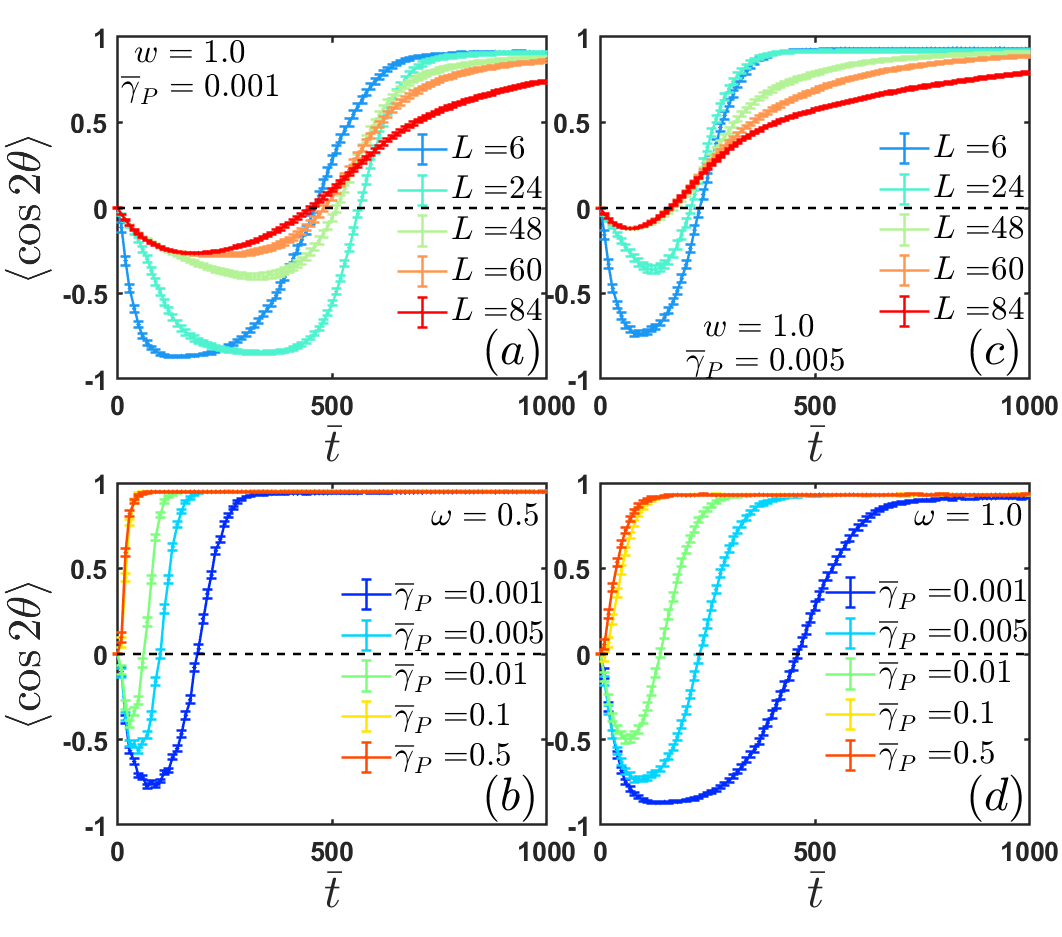}
\caption{(a) Order parameter $\langle \cos2\theta\rangle$ as a function of time for different system sizes. The phonon frequency $\overline{\omega}=1$, and the phonon damping rate $\overline{\gamma}_P = 10^{-3}$. (b) Similar as (a) but for slightly larger damping $\overline{\gamma}_P = 5\times 10^{-3}$. (c)  $\langle \cos2\theta\rangle$ as a function of time for different damping rates. The phonon frequency $\overline{\omega}=0.5$. The linear system dimension $L=6$. (d) Similar to (c) but for $\overline{\omega}=1$.}
\label{fig:3Dphonon}
\end{figure}

Fig.~\ref{fig:3Dphonon}(a) shows the time evolution of the order parameter $\langle \cos2\theta\rangle$ after a quench from high temperature disordered state to low temperature ordered state. We set phonon frequency $\overline{\omega}=1$ and a weak damping $\overline{\gamma}_P = 10^{-3}$. The initial and final temperatures are respectively $\overline{T}_i = 12$ and $\overline{T}_f = 0.1$. The coupling constant $\overline{g} = 0.2$. We observe a transient competing order as indicated by $\langle \cos2\theta\rangle<0$ at the early time. Similar behavior is also found for slightly larger phonon damping rate $\overline{\gamma}_P = 5\times 10^{-3}$ (Fig.~\ref{fig:3Dphonon}b). In particular, the life time of the transient competing order is shorter due the faster relaxation of non-equilibrium phonon fluctuations. 

Fig.~\ref{fig:3Dphonon}(d) shows a systematic study of the impact of phonon damping rate $\overline{\gamma}_P$ on the transient selection. We use a smaller system size $L=6$ to save the computational cost; nevertheless, as indicated by the panel (a), the behavior at this size is already qualitatively similar to larger size systems. We observe a transient state for slow phonon relaxation, and no clear signature of such a state for fast relaxation. We find similar behavior for a different, representative phonon frequency $\overline{\omega} = 0.5$ (Fig.~\ref{fig:3Dphonon}(c)). Crucially, the phenomenology for the underdamped phonons is highly reminiscent of the overdamped phonons.

Therefore, our numerical data provide clear, direct evidence for the robustness of our results upon accounting for the underdamped dynamics of phonons. This robustness lies in the fact that the out-of-equilibrium phonon fluctuations that select transient magnetic order are incoherent, which is very different from the more familiar Floquet engineering story, where a persistent periodic motion is crucial. Before the thermal quench, the distribution of $Q$ can be described by a Gaussian centered at $Q=0$. After the quench, this distribution evolves according to the Fokker-Planck equation. In particular, its distribution remains Gaussian but variance $\langle Q^2\rangle$ evolves in time. For overdamped phonons,  the time evolution of $\langle Q^2\rangle$ decays exponentially. For underdamped phonons, $\langle Q^2\rangle$ has an oscillatory component superimposed on the decaying background, which does not affect the selection qualitatively. 

\end{document}